\documentclass[twocolumn]{aa}
\usepackage{graphicx}
\usepackage{txfonts}
\begin{document}
   \title{To see or not to see a Bow Shock:}
   \subtitle{Identifying Bow Shocks with H$\alpha$ Allsky Surveys}
   \author{D. Brown\inst{1,2} \& D.\,J. Bomans\inst{2}}

   \institute{Astrophysics Research Institute, Liverpool John Moores University,
              Twelve Quays House,\\
              Egerton Wharf, Birkenhead CH41 1LD, UK\\
	      \email{db@astro.livjm.ac.uk}
	      \and
              Astronomical Institute, University of Bochum,
              Universit\"atsstrasse 150, D--44780 Bochum\\
              \email{dbomans@astro.rub.de}}
	
   \abstract{
OB--stars have the highest luminosities and strongest stellar winds
of all stars, which enables them to interact strongly with their surrounding
ISM, thus creating bow shocks.
These offer us an ideal opportunity to learn more about the ISM.
They were first detected and analysed around runaway OB--stars using 
the IRAS allsky survey by van Buren et al. (\cite{Buren95}).
Using the geometry of such bow shocks information concerning
the ISM density and its fluctuations can be gained from such infrared observations.
As to help to improve the bow shock models, additional observations at other wavelengths,
e.g. H$\alpha$, are most welcome.
However due to their low velocity these bow shocks have a size of $\sim 1$\degr, and
could only be observed as a whole with great difficulties.
In the light of the new H$\alpha$ allsky surveys (SHASSA/VTSS) this
is no problem any more.\\
We developed different methods to detect bow shocks, e.g. the improved determination
of their symmetry axis with radial distance profiles. Using two H$\alpha$--allsky 
surveys (SHASSA/VTSS), we searched for bow shocks and compared the different
methods. From our sample we conclude, that the correlation between the direction of both proper motion and the
symmetry axis determined with radial distance profile is the most promising detection method.\\
We found eight bow shocks around \object{HD 17505}, \object{HD 24430}, \object{HD 48099}, \object{HD 57061}, \object{HD 92206}, \object{HD 135240}, \object{HD 149757}, and \object{HD 158186} from 37 candidates
taken from van Buren et al. (\cite{Buren95}). Additionally to the traditional determination of
ISM parameters using the standoff distance of the bow shock, another approach
was chosen, using the thickness of the bow--shock layer. Both methods
lead to the same results, yielding densities ($\sim 1$\,cm$^{-3}$) and the maximal 
temperatures ($\sim 10^4$\,K), that fit
well to the up--to--date picture of the Warm Ionised Medium.
     }

   \maketitle
%

\section{Introduction}
OB--stars are the most massive and luminous stars known
with masses greater than 10\,$M_\odot$ and effective
temperatures ranging from 10\,000\,K up to 50\,000\,K.
During their short lifetime ($\leq$20$\times$10$^6$\,yr) OB--stars lose
mass at rates of 
$\dot{M}\approx10^{-7}-10^{-5}\,M_\odot$ yr$^{-1}$ (Lamers \& Cassinelli \cite{Lamers}). This stellar wind
with velocities of $v_\infty\approx 1\,000 - 3\,000$\,km\,s$^{-1}$ 
transfers a great amount of mechanical energy to the surrounding
ISM, comparable to a supernova 
explosion. As a result, OB--stars create a stellar bubble (Castor et al. \cite{castor})
which structures the Interstellar Medium (ISM). This spherical bubble is
altered when the OB--star is in motion, as wind and ISM interact directly.
The resulting nebula
is known as a bow--shock nebula. Recently many new bow shocks (size $\sim$1\arcmin) created by 
fast moving neutron stars or pulsars have been detected and
analysed.
Their geometry could be successfully used to gain information about the ISM, such
as density and temperature (Gaensler et al. \cite{gaensler}).
Thus the theoretical models of such bow shocks have become quite exact recently,
enabling us to glance at the missing link
between density fluctuations seen in \ion{H}{i} (at scales $\sim0.1-200$\,pc) and
towards pulsars (at scales $\sim5-100$\,AU).
 Bow--shock nebula around
OB--stars can also be used as ISM probes, if observable or rather detectable, as
described in the following.\\
In \cite{Buren88} van Buren \& McCray detected structures around OB--stars
and Wolf--Rayet stars
using the 60 $\mu$m allsky survey of the {\it InfraRed Astronomical Satellite} (IRAS).
These images revealed an arc--like structure and a high colour temperature, possibly being
bow shocks. A more complete sample of 188 runaway OB--stars was analysed by van Buren 
et al. (\cite{Buren95}) (hereafter VB) using the IRAS 
allsky survey which lead to the detection of 58 bow 
shocks. Because of these numerous detections and the alignment of the 
symmetry axis of the structures along the direction of proper motion of the
central star,
they could only be bow--shock nebula.\\
Due to the low resolution of IRAS ($\sim$1\farcm5)
it is difficult to determine the exact location of the bow shock and 
its symmetry axis.
Furthermore, VB could only use the Hipparcos input catalogue (HIC) to determine the
proper--motion direction of the central star. 
As the Hipparcos catalogue is now completed, its astrometrical data together with 
the almost complete H$\alpha$ allsky surveys {\it Virginia Tech Spectral Survey}$^1$ (VTSS;
Dennison et al. \cite{dennison})
and {\it Southern Hemispheric H$\alpha$ Sky Survey Atlas}\footnote{Supported 
by the National Science Foundation} (SHASSA; Gaustad et al. \cite{gaustad}), the VB
sample is reanalysed here, now using the H$\alpha$ emission line.\\
The development from a stellar bubble to a bow shock is illustrated in 
Sect. 2.
The used H$\alpha$--data and its acquisition is described in Sect. 3. Sect. 4 
explains the analysis of the bow shocks and the
improved methods developed to detect them. 
In Sect. 5 the results of the observations are given and in Sect. 6 ISM parameters derived, followed by a discussion
in Sect. 7. Finally, Sect. 8 draws a conclusion concerning the main points of this paper.

\section{Scenario}
\label{Szenario}
\begin{figure}
\centering
   \includegraphics[width=4.3cm]{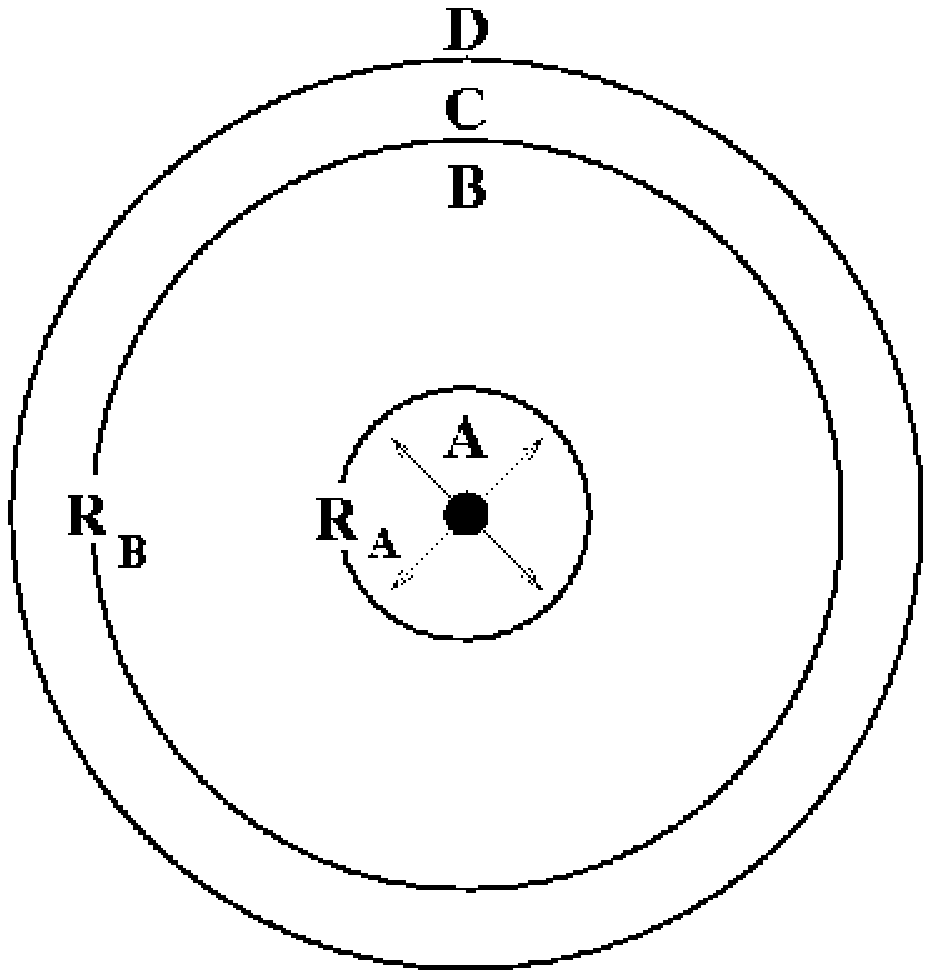}
   \includegraphics[width=4.3cm]{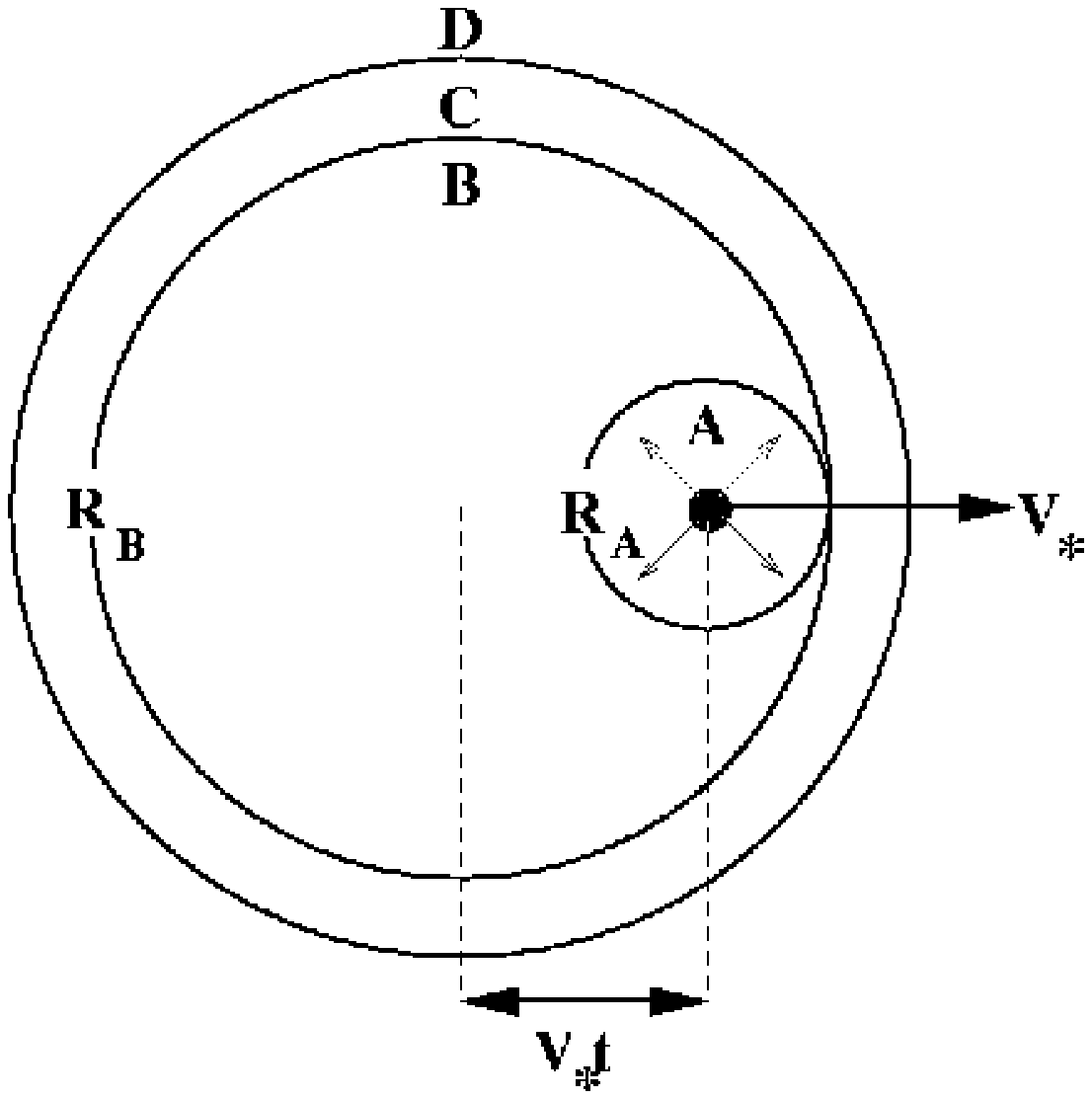}
   \caption{Structure of a stellar bubble created by a 
            stationary star (left) and a star in motion (right).
            The different regions A, B, C, and D and the geometry
            are described in Sect. \ref{Szenario}.}
      \label{Bubble}
\end{figure}

For a better understanding of the methods used, the
scenario of a bow shock and the source of the high velocity of the OB--stars
are given.\\
The evolving stellar bubble, as illustrated in Fig. \ref{Bubble}, is in its longest 
snowplow stage divided into four parts 
centred upon the OB--star (Castor et al. \cite{castor}):
The innermost area (A) is an unshocked and freely expanding
stellar wind, followed by a larger area (B) of shocked stellar
wind.
These hot regions, as if they were a snowplow (hence the name of this stage),
have pushed together a thinner area (C) of shocked ISM. All regions 
are embedded in the unshocked ISM (D).\\ 
The models describing such stellar bubbles assume that the OB--star is
stationary with respect to the ISM.
However as all stars have a proper motion, so do OB--stars.
A special population of OB--stars is known as runaway OB--stars (Blaauw
\cite{Blaauw}).
They are defined as
having a proper motion greater than 30\,km\,s$^{-1}$. This criteria was chosen
to
discern them from non--runaway OB--stars, which have a velocity dispersion
of about 10\,km\,s$^{-1}$.\\
As runaway stars are often found in isolated regions,
their
high velocity cannot be explained as a motion within a stellar cluster.
Two scenarios explaining observed properties of runaway OB--stars are favoured:\\
Firstly, the Binary--Supernova--Scenario (BSS) as described by Blaauw
(\cite{Blaauw}).
The partner of the OB--star explodes as a Supernova (SN). Thereby the OB--star is set free
with its typical orbital velocity of 30--150\,km\,s$^{-1}$.
And secondly, the Dynamical--Ejection--Scenario (DES) proposed by Hoffer
(\cite{Hoffer}).
In this Scenario the collision of two binary systems leads to the ejection of
one star with a velocity of up to 200\,km\,s$^{-1}$.\\
\begin{figure*}
\centering
   \includegraphics[width=8.9cm]{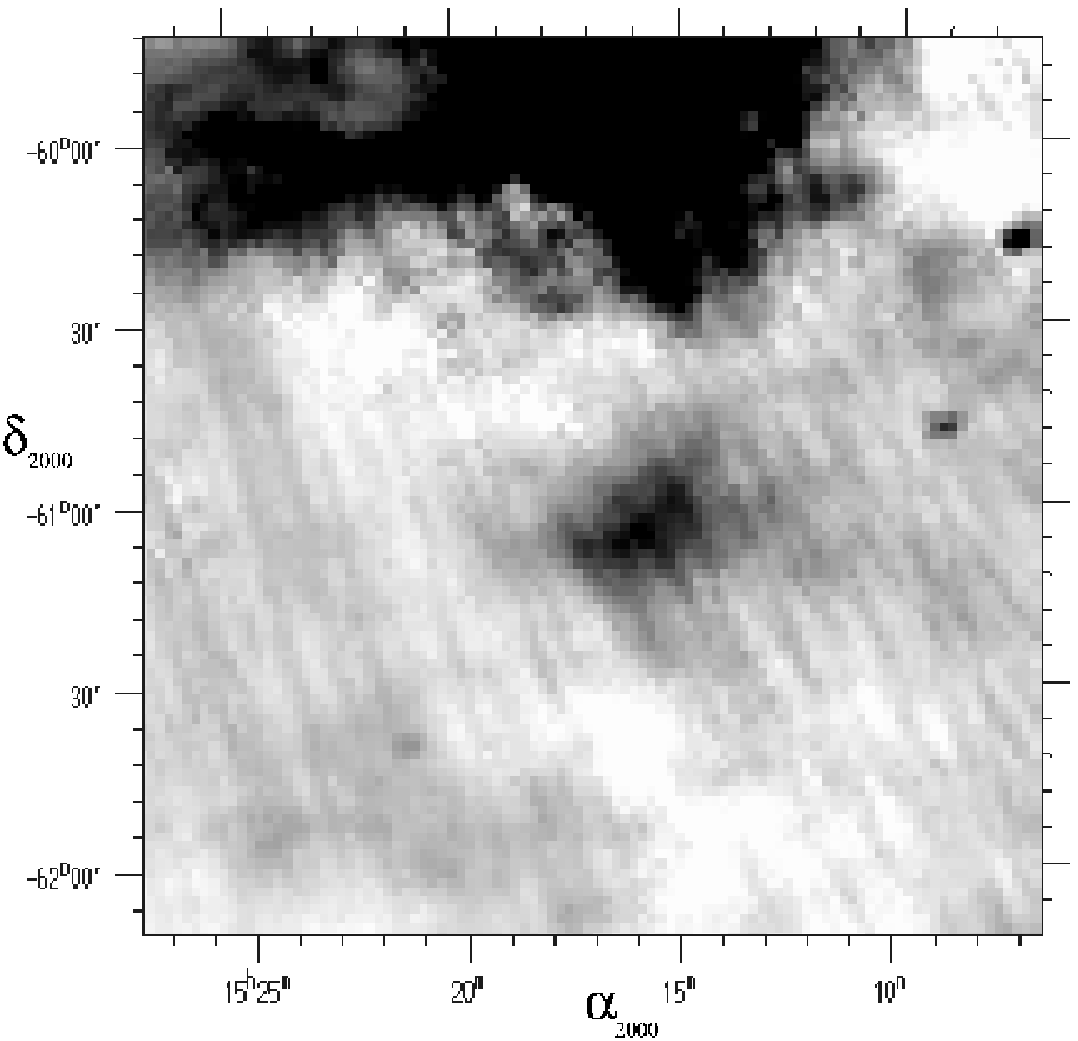}
   \includegraphics[width=8.9cm]{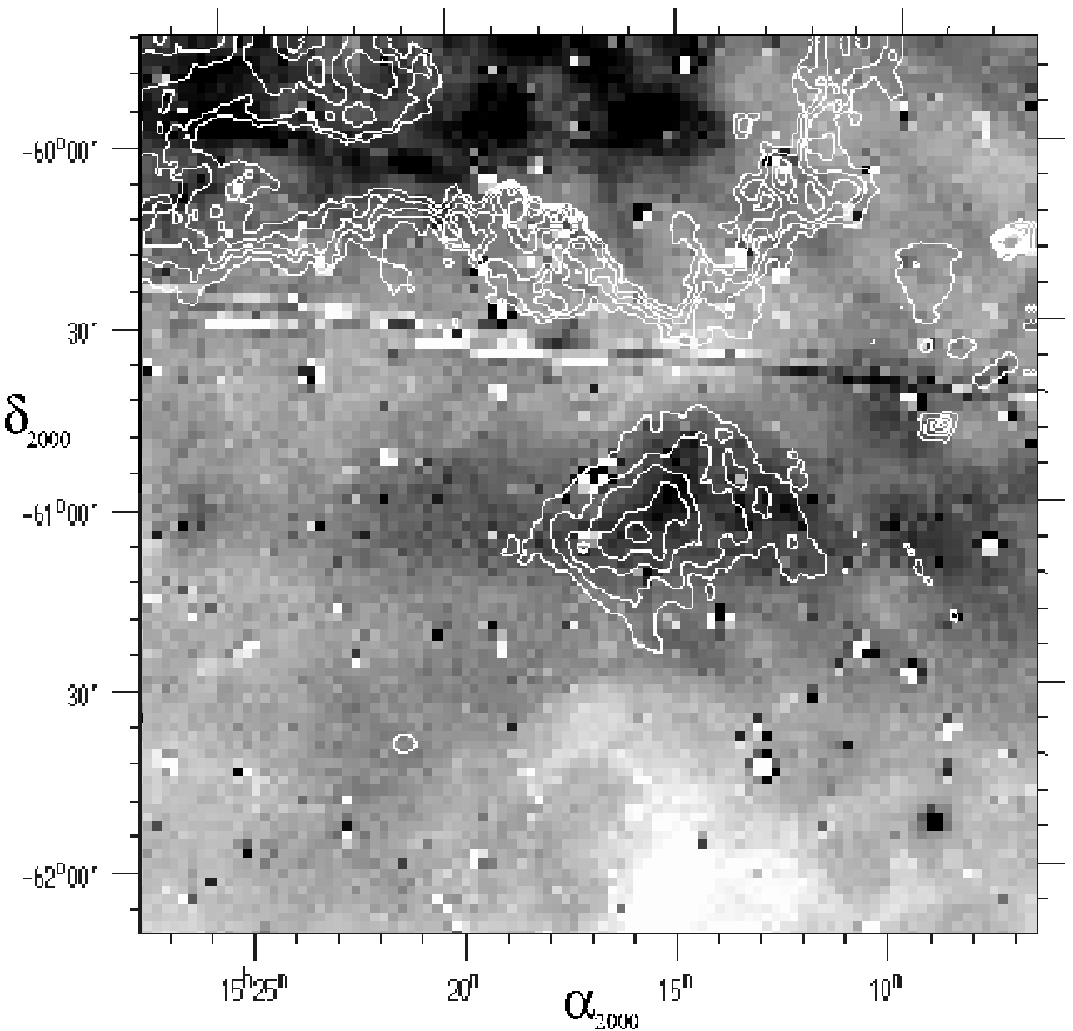}
   \caption{Left: IRAS 60\,$\mu$m excess map of \object{HD 135240}. Right: 
            Overlay of H$\alpha$ image in grey--scale and the IRAS 60\,$\mu$m excess map in
            white contours. Both are shown with their slanted
            equatorial--coordinate system (J2000) and inverted grey--scale.}
      \label{Overlay}
\end{figure*}
Taking the motion of the star ($V_\ast$) into account, regions A, B, and C will still be
spherical as long as the stellar velocity is smaller than the sound velocity 
within B. However B and C are no longer centred on the star, as shown in Fig \ref{Bubble}. If region A
and D do not interact directly, this would be
the only alteration to the model of the stellar bubble.
But as soon as the star enters denser regions of the ISM, like molecular
clouds, the cooling of region B becomes more effective. This leads to a
collapse of B and C within timescales smaller than the lifetime of the
OB--star, 
and the approach of region A and D. As molecular clouds are 
not frequently encountered, region A and D can only interact when the offset
of A 
to B and C is $V_\ast t\approx R_B(t)-R_A(t)$. Where the timescale is dependent upon
the velocity of the OB--star and the density of the surrounding ISM. Taking
typical values of the lifetime and velocity $V_\ast$=39\,km\,s$^{-1}$ 
this leads to a density constraint of $n\ge 0.017$\,cm$^{-3}$ in which 
A and D can interact directly.\\
In the case of directly interacting unshocked stellar wind and unshocked ISM,
the geometry is changed completely. As Wilkin (\cite{Wilkin}) describes, the
ram pressure
of both media can be balanced directly and result in a bow shock. This bow
shock
is axi--symmetric along the direction of proper motion and can be approximated
by a parabola. The two layers B and C of the model above are mixed due to
turbulence
and plasma instabilities leading to a single layer in which the material of the ISM
and stellar wind moves along the bow shock. The material in this has
experienced
a nearly isothermal shock, so its density is higher than that of the 
surrounding ISM. This leads to the creation of warm
interstellar dust best seen in 60\,$\mu$m, and the OB--star in the centre leads 
to the ionisation of the layer emitting H$\alpha$.

\section{Data}
\subsection{Selection}
\label{Data}
The data used for this program were taken from the SHASSA and incomplete VTSS allsky
surveys. SHASSA
contains the southern hemisphere up to $\delta$=15\degr~and
VTSS the northern hemisphere down to $\delta=$--15\degr.
Both surveys were made with a CCD detector and a fast photo--objective of
$\sim$55\,mm at $\sim$f/1.4, leading to a field of view of $\sim$13\degr. All
images were integrated $\sim$25\,min resulting in a detection limit down
to $\sim$0.75 rayleigh $\left ( 1\,{\rm R}=\frac{10^6}{2\pi}\frac{\rm
Photons}{\rm cm^2\,\rm s\,\rm sr} \right )$.
Due to the different pixel sizes of the detectors SHASSA has a resolution of
0\farcm8
and VTSS of 1\farcm6.\\
The H$\alpha$--sample used consists of the O--stars taken from the VB
sample with data from either VTSS or
SHASSA, ensuring a sufficient
Lyman continuum flux to ionise the bow--shock layer.\\
We searched for bow shocks around these 37 candidates of the H$\alpha$--sample
within the SHASSA and VTSS
H$\alpha$ survey. Due to always present background nebulosity, it had to be
ensured
that the structures seen were really bow shocks. As described
at the end of Sect. \ref{Szenario}, the bow--shock layer 
should be visible in 60\,$\mu$m as well as in the H$\alpha$ emission line. Therefore,
the H$\alpha$
images were compared with the 60\,$\mu$m IRAS images of the same region. Though
the IRAS images show a great amount of nebulous emission, the nebulosity can successfully be subtracted
using the 
100\,$\mu$m images of IRAS. This correction has been done based on the recipe of
VB with the creation
of IRAS 60\,$\mu$m excess maps, shown for the example of \object{HD 135240} in Fig. \ref{Overlay}.\\
To compare both images with each other, the contours of the IRAS 60\,$\mu$m excess map were
overlaid upon
the H$\alpha$ images (see Fig. \ref{Overlay}). We used this image to decide whether a
bow shock seen in
the IRAS 60\,$\mu$m excess map is also present within the H$\alpha$ image. The eight bow shock
detections and a short description of the comparison using the overlays are 
given in Table \ref{halphatable}. The IRAS 60\,$\mu$m excess images and their corresponding
H$\alpha$ images are also shown in Fig. \ref{HalphaBilder1} 
and Fig. \ref{HalphaBilder2}.\\
\begin{figure}
\centering
   \includegraphics[width=8cm]{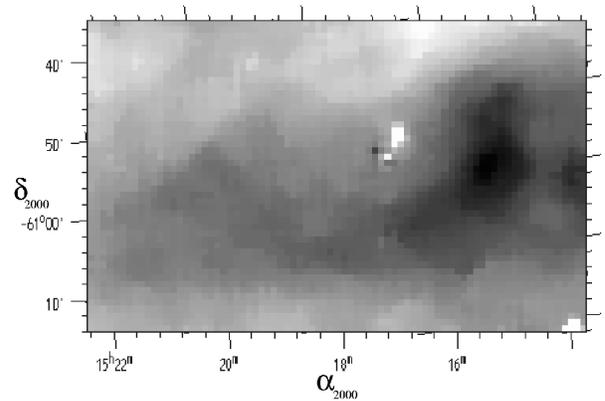}
   \caption{Median filtered section of the H$\alpha$ image containing 
            the bow shock around \object{HD 135240}. The coordinate system and grey--scale as in Fig. \ref{Overlay}.}
      \label{Median}
\end{figure}
To check for the positioning quality of the H$\alpha$ image compared 
to the IRAS 60\,$\mu$m excess maps in the overlays, 
we could not use the nebulosities as criteria. As we are using these overlays
to see
coinciding positions of nebulosities within both images this would be
misleading. 
Better criteria are point 
sources like stars which have to be visible with IRAS and in VTSS/SHASSA, thus, we used 
the positions of M giants as a reference. Using this 
method, a deviation of the positional offsets of $\sigma=20$\arcsec~was measured, 
which is sufficiently smaller 
than the resolution of both images.\\
To analyse the H$\alpha$ images of the selected bow shocks 
only the interesting region was extracted.
As only the search for bow shocks using the overlays requires
the best possible resolution of 1.6\arcmin~(VTSS) and 0.8\arcmin~(SHASSA),
the images could for further analysis be median filtered with a 
$5\times5$ pixel wide box (see Fig. \ref{Median} for the example of \object{HD 135240}). The
resulting improvement
of S/N leads to a decreased resolution of 8\arcmin~(VTSS) and 4\arcmin~(SHASSA).
   \begin{table}
      \caption[]{Surveys containing the candidates of the H$\alpha$--sample;
bow--shock detections as results from the overlay are printed in bold letters.}
         \label{halphatable}
\vspace*{-0.7cm}
     $$ 
         \begin{array}{p{0.2\linewidth}p{0.25\linewidth}p{0.4\linewidth}}
            \hline
            \noalign{\smallskip}
            Star& Survey& Bow Shock\\
            \noalign{\smallskip}
            \hline
            \noalign{\smallskip}
 HD 1337&    VTSS&   ambiguous\\
 \bf HD 17505&   \bf VTSS&   \bf complicated background\\
 HD 19820&   VTSS&   weak\\ 
 \bf HD 24431&   \bf VTSS&   \bf detection\\
 HD 30614&   {\sl (VTSS)}&     {\sl not in survey}\\
 HD 34078&   {\sl (VTSS)}&     {\sl not in survey}\\
 BD +39 1328& VTSS&  non--detection\\
 HD 37020&   VTSS/SHASSA& ambiguous\\
 HD 41161&   {\sl (VTSS)}&     {\sl not in survey}\\
 HD 41997&   VTSS&   ambiguous\\
 HD 47839&   VTSS&   non--detection\\
 \bf HD 48099&   \bf VTSS/SHASSA& \bf complicated background\\
 HD 52533&   SHASSA&  non--detection\\
 HD 54662&   VTSS/SHASSA& ambiguous\\
 \bf HD 57061&   \bf SHASSA&  \bf detection\\
 HD 64315&   SHASSA&  ambiguous\\
 HD 66811&   SHASSA&  non--detection\\
 \bf HD 92206&   \bf SHASSA&  \bf small\\
 HD 101131&  SHASSA&  non--detection\\
 HD 112244&  SHASSA&  non--detection\\
 HD 130298&  SHASSA&  ambiguous\\
 \bf HD 135240&  \bf SHASSA&  \bf detection\\
 HD 329905&  SHASSA&  non--detection\\
 \bf HD 149757&  \bf SHASSA&  \bf detection\\
 HD 156212&  SHASSA&  ambiguous\\
 \bf HD 158186&  \bf SHASSA&  \bf complicated background\\
 HD 164492&  SHASSA&  non--detection\\
 HD 169582&  SHASSA&  ambiguous\\
 HD 175514&  VTSS&    weak\\ 
 HD 186980&  {\sl (VTSS)}&      {\sl not in survey}\\
 HD 188001&  VTSS&    non--detection\\
 HD 227018&  {\sl (VTSS)}&      {\sl not in survey}\\
 HD 195592&  {\sl (VTSS)}&      {\sl not in survey}\\
 HD 199579&  VTSS&    incomplete scan\\
 HD 203064&  VTSS&    non--detection\\
 HD 210839&  {\sl (VTSS)}&      {\sl not in survey}\\
 HD 214680&  VTSS&    non--detection\\
            \noalign{\smallskip}
            \hline
         \end{array}
     $$ 
\vspace*{-0.5cm}
   \end{table}
\subsection{Distances}
\label{result_photo}
To convert angular sizes into linear sizes, the distances of the eight
bow--shock 
candidates had to be determined. Additionally, the interstellar absorption
had to be calculated to fit the brightness profile correctly.\\
Due to the high parallax errors determined by Hipparcos for the candidates
the distances were determined using their spectral parallax with
absolute magnitudes derived from Landolt--B\"ornstein (\cite{Landolt}) according
to the spectral classification described in appendix \ref{Photometric}. As for the absorption along the line of sight,
photometric data for the B and V filters were taken, and the
normal extinction law (Mathis, \cite{Mathis}) applied. The expected absorption within the 
H$\alpha$--line was estimated according 
to the interstellar extinction given by Mathis (\cite{Mathis}).\\
In appendix \ref{Photometric} the  magnitudes 
and multiplicity of the different sources are described.
All data and results are given in Table \ref{photo_astro_data}.
   \begin{table*}
      \caption[]{Photometry and astrometry of the central stars of the bow--shock candidates. $m_B$ and $m_V$ are the
                 apparent B and V magnitudes, $\mu_{\alpha}$ and $\mu_{\delta}$ give the proper motion along the 
		 rightascension and declination axis, and $V_r$ is the radial velocity. The spatial velocity $V$,
                 the symmetry axis position $\theta_a$ and its inclination $\iota_\mathrm{A}$ are derived from the astrometric data.
                 The distance $r$ is derived from the spectral parallax.}
         \label{photo_astro_data}
\vspace*{-0.5cm}
     $$ 
         \begin{array}{p{0.1\linewidth}p{0.08\linewidth}p{0.1\linewidth}llr@{\pm}l r@{\pm}l r@{\pm}l l r@{\pm}l r@{\pm}l l}
            \hline
            \noalign{\smallskip}
            star& SC& multiplicity &m_B &m_V &\multicolumn{2}{c}{\mu_\alpha}&   \multicolumn{2}{c}{\mu_\delta}& \multicolumn{2}{c}{V_r}& V 
            &\multicolumn{2}{c}{r}&\multicolumn{2}{c}{\theta_a}& \iota_\mathrm{A} \\
            & & &[{\rm mag}] &[{\rm mag}] &\multicolumn{2}{c}{[{\rm mas\,yr}^{-1}]}&   \multicolumn{2}{c}{[{\rm mas\,yr}^{-1}]}& \multicolumn{2}{c}{{\rm km\,s}^{-1}}& {\rm km\,s}^{-1}
            &\multicolumn{2}{c}{[{\rm pc}]}&\multicolumn{2}{c}{[\degr]}& [\degr] \\            \noalign{\smallskip}
            \hline
            \noalign{\smallskip}
 HD 24431&  O9 IV--V& binary&         7.2& 6.9&    -0.21&1.01&   -1.52&0.86&   -9.9&2&   12.0&   911&57&     262&38&   56\\   
 HD 48099&  O7 V&     binary&         6.3& 6.4&     +0.81&0.65&    +2.35&0.53&    +31&2&    37.7&1829&77&     71&15&   55\\   
 HD 57061&  O9 II&    quintuple&      6.0& 6.1&    -1.82&0.44&    +3.74&0.59&    +40.4&2&  55.8& 1914&79&   116.0&6.5&  46\\  
 HD 92206&  O6 ?&     binary&         8.3& 8.2&    -10.4&4.2&     +6.8&4.2&     -10&5&    40.5&  4049&109& 147&19&       1\\
 HD 135240& O7 III-V& triple&         5.0& 5.1&    -2.02&0.51&   -4.08&0.55&     +9.2&2&   32.5& 1131&63&   243.7&6.5&  16\\  
 HD 149757& O9 V&     single&         2.6& 2.6&     +13.07&0.86&  +25.44&0.71&   -15&10&   26.8& 163&27&     62.8&1.6&  34\\     
 HD 158186& O9.5 V&   binary&         7.0& 7.0&     +1.36&1.11&   -1.26&0.48&   \multicolumn{2}{c}{-9}&  13.9&1094&62&    317&26&  41\\   
 HD 17505&  O6.5 V&   binary?&        7.8& 7.4&    -1.38&1.16&   -0.69&1.08&   -17&5&       25.7&1416&69&    207&41&  41\\   
            \noalign{\smallskip}
            \hline
         \end{array}
     $$ 
   \end{table*}
\subsection{Motion}
The proper motion of the eight bow--shock candidates was taken from the 
Hipparcos--catalogue, as well as their errors derived from the given error--ellipse.
The radial velocity information was taken from CDS (Evans D. S. \cite{evans}; Wilson R. E. \cite{wilson}).
Only in the case of \object{HD 158186} was this value updated with 
respect to the ones given in VB.
The astrometric data led to the determination of the inclination
and the position angle of the proper motion concerning the central stars of
the bow--shock candidates. The astrometric results are given in Table \ref{photo_astro_data} .\\
If the nebula were created by a bow shock,
these parameters should be the same as for the bow shocks. The inclination
can be directly compared. 
As for possible image rotation of the H$\alpha$ image compared to the global
coordinate system, neighbouring stars were measured and the position angle corrected.

\section{Analysis}
\begin{figure}
\centering
   \includegraphics[width=8cm]{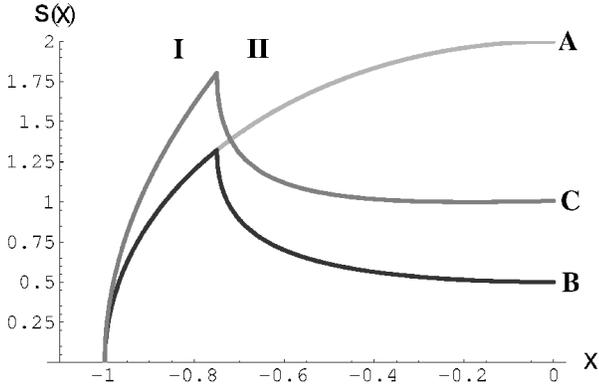}
   \caption{Theoretical radial brightness plot for a sphere (A) light grey, 
            a spherical shell (B) dark grey and a bow--shock layer (C) medium grey.
            They are derived from the line of sight $s(x)$ through the regions,
            with the star at the position $x=0$. I and II are
            the exterior and interior of the different regions.}
      \label{RBPSkizze}
\end{figure}

\begin{figure*}
\centering
   \includegraphics[width=8.9cm]{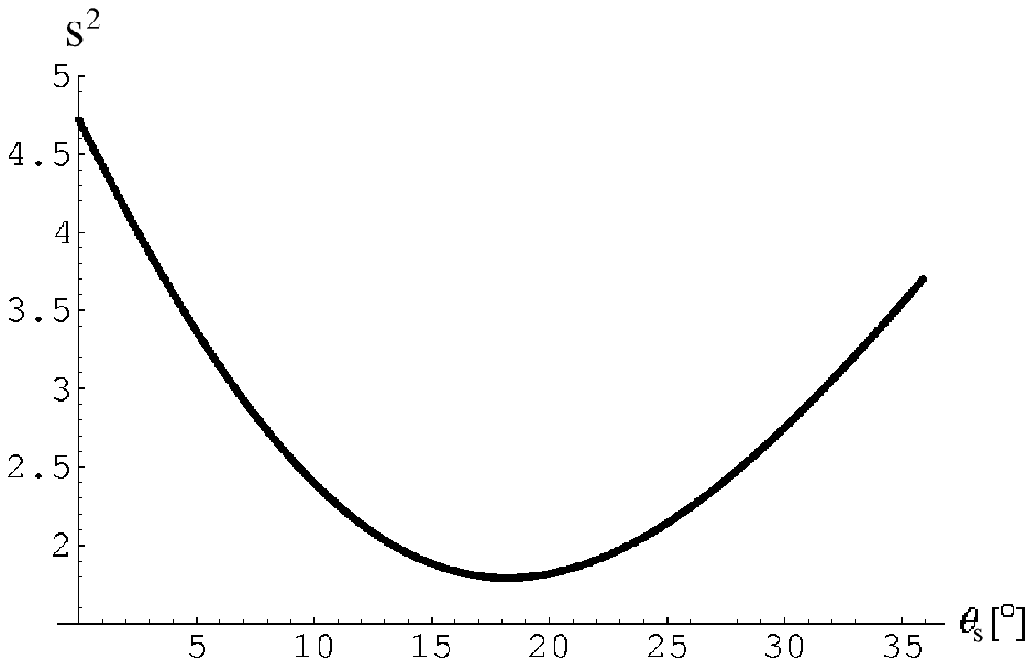}
   \includegraphics[width=8.9cm]{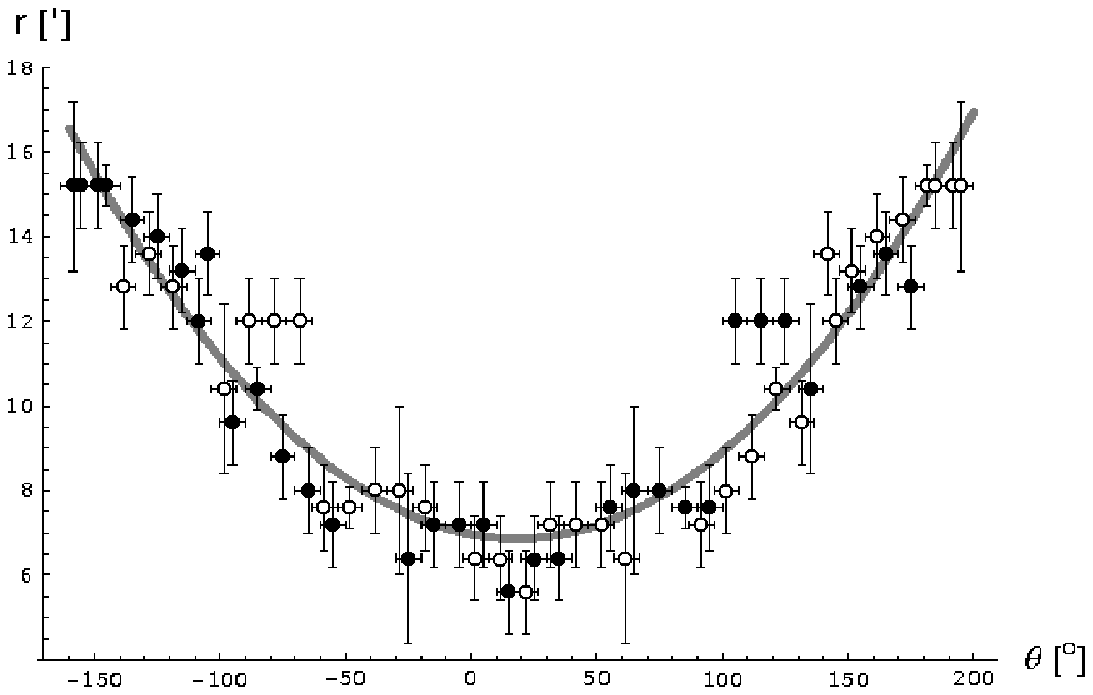}
   \caption{Left: The variance $s^2$ of different symmetry axis $\theta_{\rm s}$ referring to an arbitrary axis for \object{HD 158186}. Right: The radial distance profile of \object{HD 158186} with the position angle 
            $\theta$ as used in the left figure. Symmetry axis is chosen as the minimum
            of $s^2$. Open circles indicate the mirrored data.}
      \label{RDP}
\end{figure*}

\begin{figure}[h]
\centering
   \includegraphics[width=8.8cm]{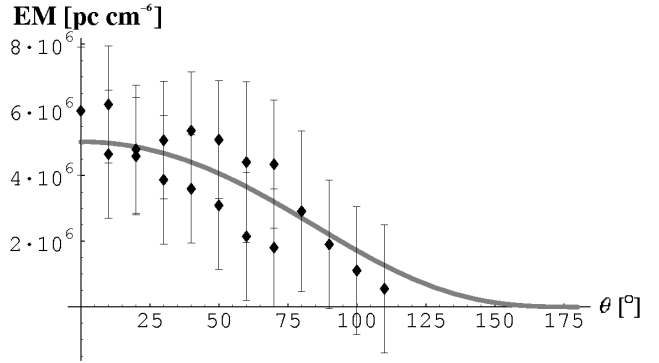}
   \caption{The brightness profile of \object{HD 57061} resulting in a thickness $d=(16.9\pm4.4)\times10^{15}$\,cm.}
      \label{BPSkizze}
\end{figure}

\begin{figure*}
\centering
   \includegraphics[angle=0,width=8.9cm]{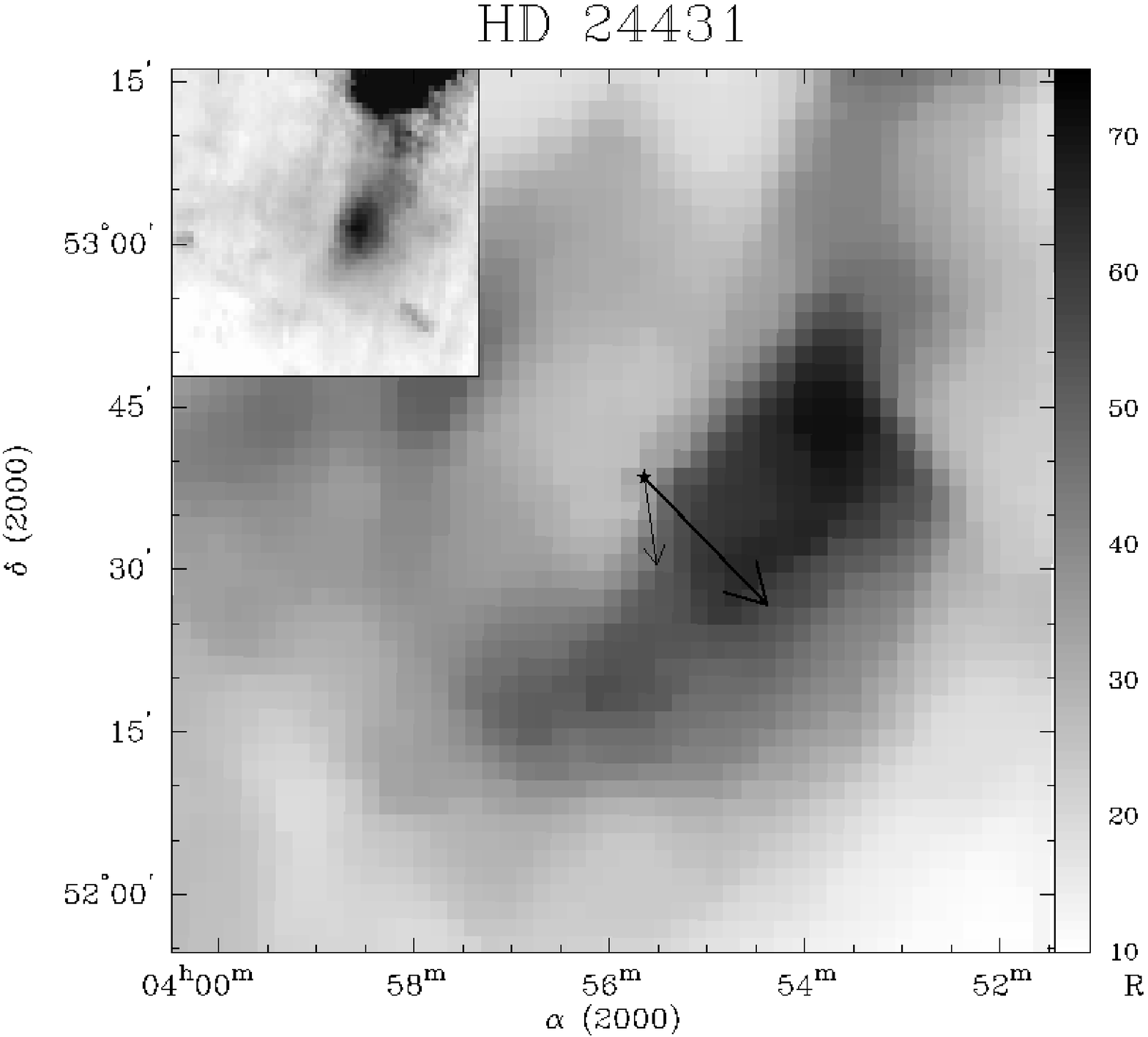}
   \includegraphics[angle=0,width=8.9cm]{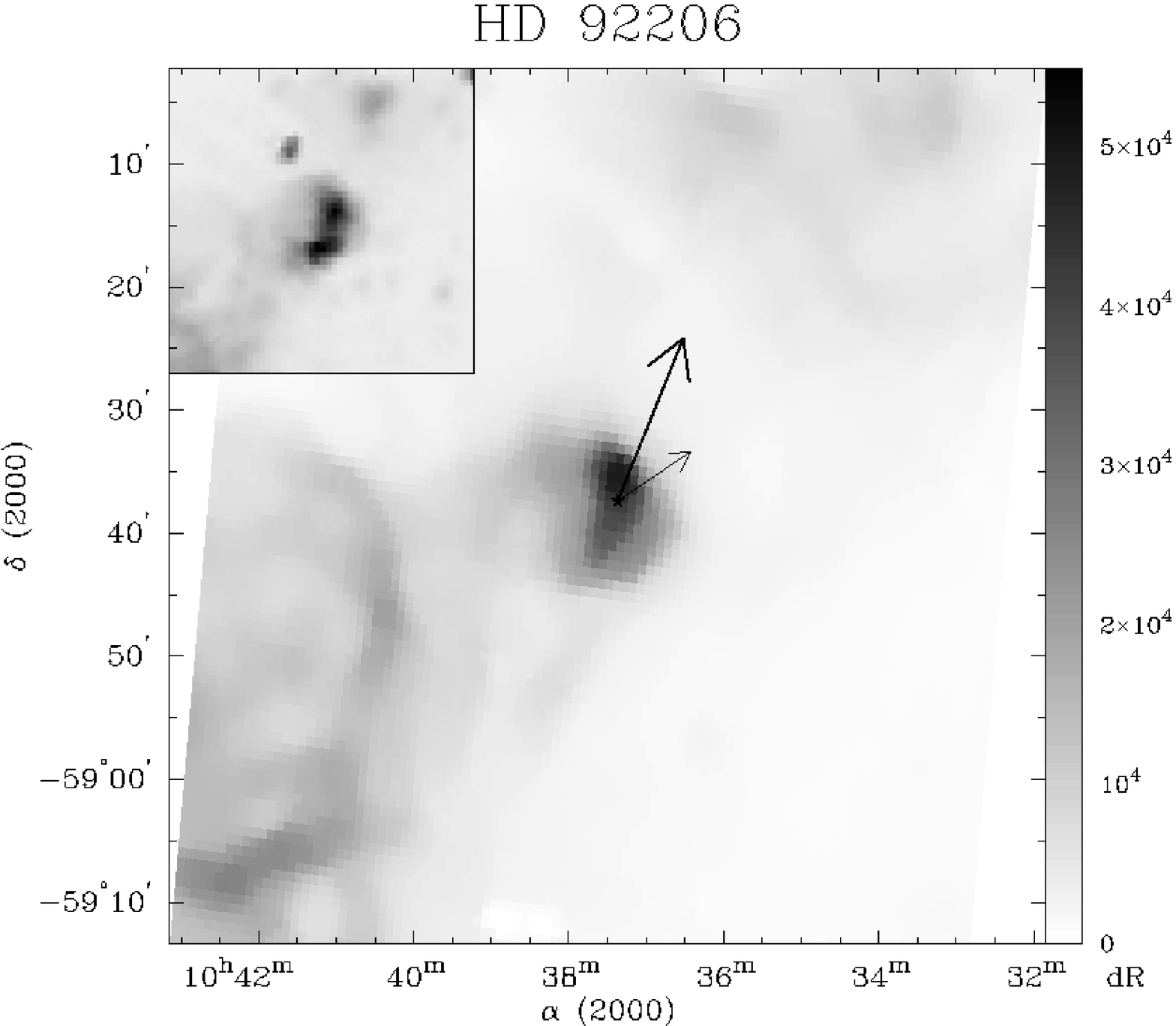}
   \includegraphics[angle=0,width=8.9cm]{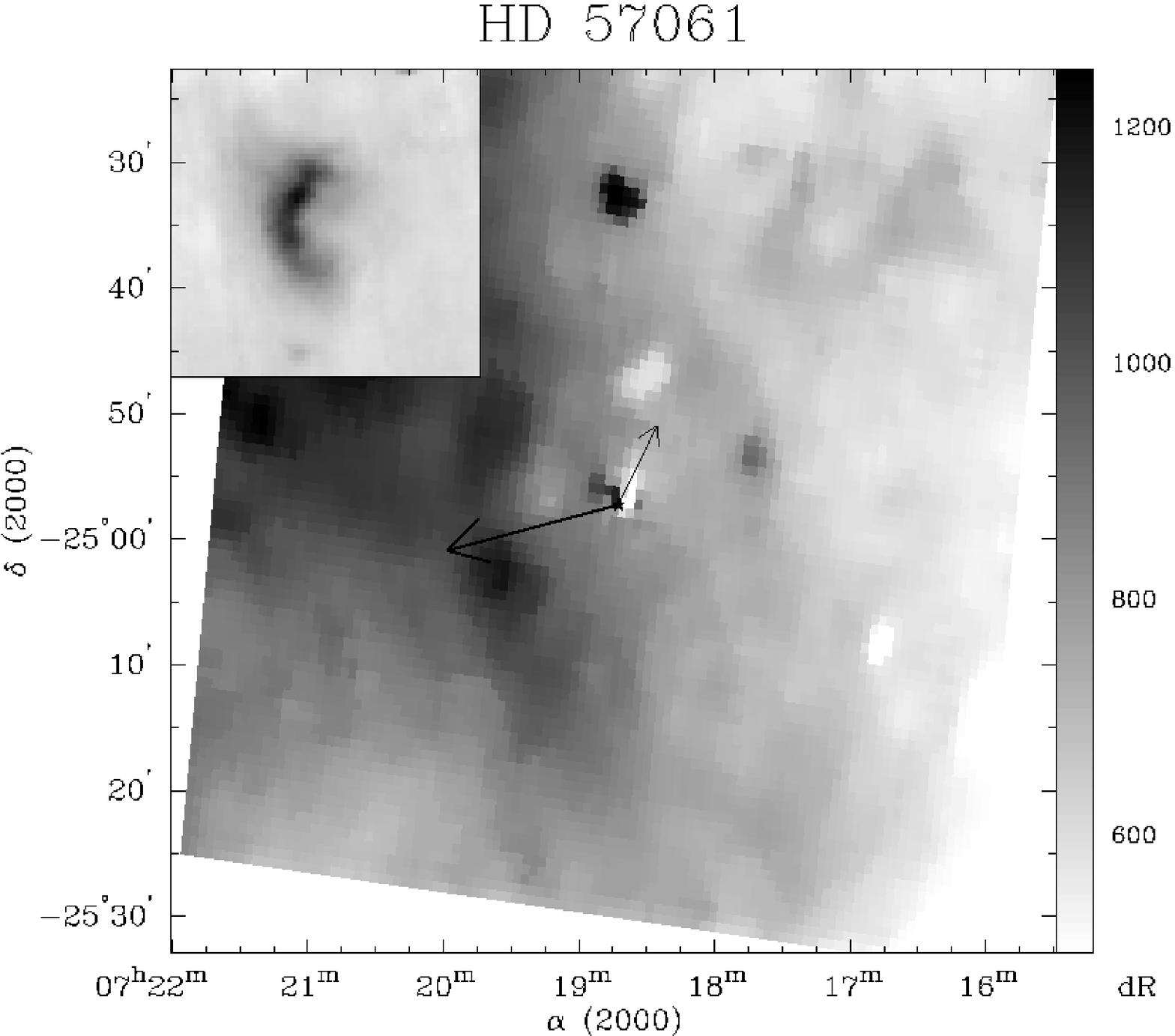}
   \includegraphics[angle=0,width=8.9cm]{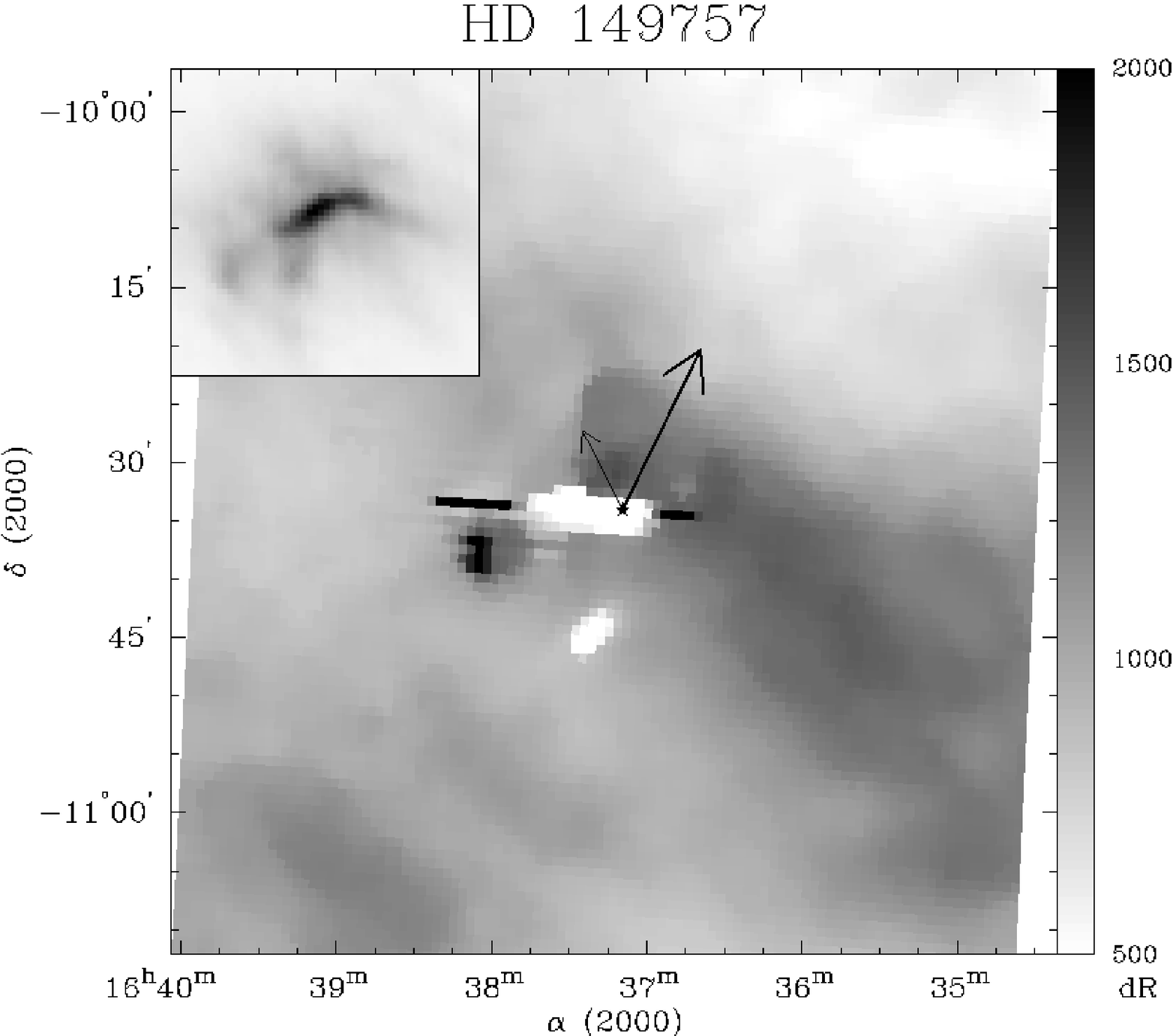}
   \caption{Median filtered H$\alpha$ images of the first four
            bow--shock candidates. The short arrow indicates the proper
            motion direction of the star and the long arrow the position
            of the symmetry axis. The position of the star is given
            by a star symbol. The grey--scales used are given in rayleigh 
	    (R) or decirayleigh (dR). The inset shows the IRAS 60\,$\mu$m excess images of the 
            bow--shock candidates used to determine the overlays,
            their size scaled to the H$\alpha$ images
	    and using inverted grey--scales.}
      \label{HalphaBilder1}
\end{figure*}

Bow--shock nebulae are structures which appear limb brightened, due to their
shell like geometry. Assuming that the gas within the layer is optically thin, one
can
compute qualitatively the characteristics of a brightness profile by
determining
the length of the line of sight within the layer. When
plotting this profile
against the radial distance of the line of sight, one obtains radial
brightness plots 
as shown in Fig. \ref{RBPSkizze}. Case A demonstrates the situation of limb darkening
for a sphere, while
B shows the radial brightness plot of a spherical shell. The radial brightness plot of a bow
shock is that
of case C. The inner boundary of the layer was described by a parabola, as van
Buren et al. (\cite{Buren90})
suggested. The outer boundary was described by a confocal parabola ensuring a
constant thickness
of the layer. The similar appearance of radial brightness plot B and C will be discussed later.\\
\begin{figure}
   \includegraphics[width=8.5cm]{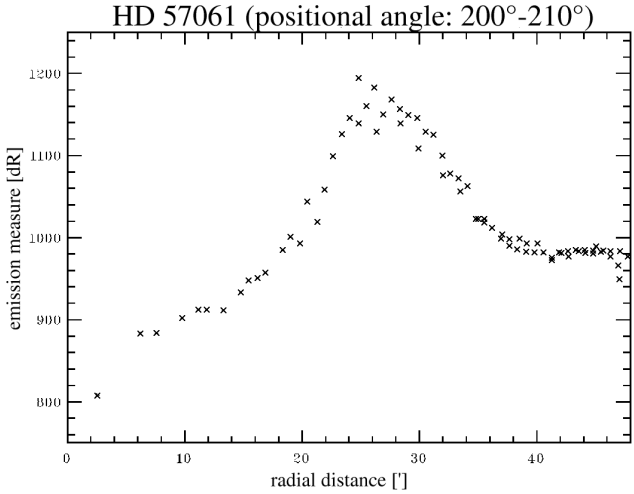}
   \caption{The measured radial brightness plot of \object{HD 57061} with the radial distance from the
            central O--star.}
      \label{BPplot}
\end{figure}
For all eight candidates, detected using IRAS 60\,$\mu$m excess maps (see Sect. \ref{Data}), radial brightness plots were 
derived by transforming the pixel
coordinates of the
images into a polar coordinate system centred upon the central O--star. The
surface H$\alpha$ brightness of
all pixels within a 10\degr~wide wedge were then plotted against their radial
distance in
arcmin. A resulting radial brightness plot for \object{HD 57061} is shown in Fig. \ref{BPplot},
demonstrating the typical limb brightening
of a bow shock superimposed upon a nearly constant background emission.\\
The measurement of the symmetry axis of the bow--shock nebula was verified with
a method, other than that
proposed by VB. When applying their method one determines the symmetry axis of
the structure
created by the bow shock and the background emission {\it together}. This would
alter the direction of
the bow--shocks symmetry axis. If one could determine the location of the inner
boundary of the bow--shock 
layer free of background contaminations, it would be possible to determine the symmetry axis
of the
bow shock by itself. This can be achieved using the location of the
maximum of the radial brightness plot.
The maximum is determined by the varying
brightness of the bow shock alone. Background nebulosity will most certainly
not be so sharply
peaked. Therefore, its location exactly traces the inner boundary.\\
The radial distance of the inner boundary was traced with radial brightness plots for each bow--shock
candidate. Plotting the distance
against the position angle of the wedge used to create the radial brightness plot, results in a
radial distance
profile as shown on the right of Fig. \ref{RDP} for the example of \object{HD 158186}. 
The radial distance profile shows a symmetric
behaviour and its symmetry axis coincides
with that of the bow--shock structure. Taking the existence of a symmetry
axis for granted the position of the axis is determined by a symmetric
function fitted
to the data points, which also coincides with the symmetry axis
of the bow shock. A parabola:
\begin{equation}
\label{fitparabel}
r=a(\theta-\theta_s)^2+b
\end{equation}
was chosen to fit the data because of its few free parameters and its close
approximation of the data.
The parabola is fitted to the data plotted in polar coordinates.
A parabola in cartesian coordinates, for which the parabola of 
van Buren et al. (\cite{Buren90}) in Eq. \ref{burenparabel} is defined, can be transformed, but would be of a more complicated
structure not
needed to measure the symmetry axis. Resulting from the great width of the
wedge, only few radial brightness plots were created, thus only
few data points were present in the radial distance profiles, but
enough to ensure a symmetric distribution (see open or filled 
circles on the right of Fig. \ref{RDP}). To double the data points 
and gain a more precise symmetry axis position the points were mirrored
with an assumed
symmetry axis $\theta_s$. Thereafter, the parabola (Eq. \ref{fitparabel})
was fitted to the points, using
the least--square method and error weighting (as for all
following fits), keeping the displacement
$\theta_s$ constant.
In steps of 0\fdg1, $\theta_s$ was changed and an variance $s^2$ of the
fit derived.
The symmetry axis was chosen to be at a $\theta_s$ of minimal
variance (see the left of Fig. \ref{RDP} for the example of \object{HD 158186}), 
which agrees with the less precise value found without mirroring.
The errors of $\theta_s$ result
from a final fit allowing for $\theta_s$ to vary.
All symmetry axes are given in Table \ref{BSparameters}.\\
The comparison of all position angles determined either by VB or through
radial distance profile or astrometry are given in Table \ref{compairsymaxis}, also noting the deviation of radial distance profile
with respect to the astrometric results.
   \begin{table}
      \caption[]{Derived parameters of the analysed bow shocks. $\theta_{\rm s}$ and $\iota$ are the position and inclination
                 of the symmetry axis using the H$\alpha$ images. The standoff distances SOD are given for the two cases of
		 no inclination and inclination. d is the thickness of the bow--shock layer perpendicular to its surface.
                  Uncertain values are given without errors.}
         \label{BSparameters}
\vspace*{-0.5cm}
     $$ 
         \begin{array}{p{0.18\linewidth}r@{\pm}lr@{\pm}lr@{\pm}lr@{\pm}lr@{\pm}l}
            \hline
            \noalign{\smallskip}
            star&\multicolumn{2}{c}{\theta_{\rm s}}  &\multicolumn{2}{c}{SOD
            (\iota=0\degr)}&\multicolumn{2}{c}{SOD (\iota)}&\multicolumn{2}{c}{\iota}&\multicolumn{2}{c}{d}\\
            &\multicolumn{2}{c}{[\degr]}  &\multicolumn{2}{c}{[\arcmin]}&\multicolumn{2}{c}{[\arcmin]}&\multicolumn{2}{c}{[\degr]}&\multicolumn{2}{c}{[10^{15}\,{\rm cm}]}\\
            \noalign{\smallskip}
            \hline
            \noalign{\smallskip}
 HD 57061 &  202.8 & 2.0&     13.56 & 0.16&    6 & 11&        75 & 40&    16.9 & 4.4\\
 HD 92206&   79.9 & 4.1&      3.11 & 0.16&  3.9 & 1.8&        0 & 10&      9.8 & 2.7^1\\
 HD 158186& 238.3 & 1.7&      5.96 & 0.13&  7.4 & 1.4&        \multicolumn{2}{c}{0}&      1.80 & 0.11\\
 HD 135240& 296.7 & 1.7&     10.65 & 0.16& 13.3 & 1.6&        \multicolumn{2}{c}{0}&      2.90 & 0.42\\
 HD 149757&  67.1 & 1.7&      9.42 & 0.18&    0 & 48&        \multicolumn{2}{c}{90}&    \multicolumn{2}{c}{\cdots}\\
 HD 17505&  358.1 & 3.8&      5.82 & 0.18&  5.8 & 2.1&        \multicolumn{2}{c}{0}&    0.59 & 0.07\\
 HD 24431&   312  & 3.1&      7.10 & 0.32&  5.6 & 13&       71 & 49&     6.6 & 1.0\\
 HD 48099&   238.6 & 5.3&     4.25 & 0.25&  \multicolumn{2}{c}{0}&         90& 0.8&     2.25 & 0.62\\
            \noalign{\smallskip}
            \hline
         \end{array}
     $$ 
\begin{list}{}{}
\item[$^{\mathrm{1}}$] Layer thickness of \object{HD 92206} is given in $10^{9}$\,cm
\end{list}
\vspace*{-0.5cm}
\end{table}
The data is transformed back to a cartesian
coordinate 
system with the x--axis as symmetry axis. Now, the physically
motivated parabola
of van Buren et al. (\cite{Buren90})
\begin{equation}
\label{burenparabel}
z=\frac{y^2}{3R_0}-R_0
\end{equation}
can be fitted to the data of $\theta=[-90\degr,+90\degr]$ nearest the apex.
The parameter $R_0$
so determined is known as the standoff distance, and is used to calculate
ISM densities.
Eq. \ref{burenparabel} is only correct for an inclination of $\iota=0$ and
thus,
$R_0$ represents an upper limit. 
The rotated parabola 
$\left ( z(x,y)=\frac{x^2+y^2}{3R_0}-R_0 \right )$ with a different inclination
angle $\iota$
 is more complex:
\begin{equation}
\label{kippparabel}
z(y)=\frac{y^2}{3R_0}\cos \iota - R_0 \left( \frac{3R_0}{4}\tan^2 \iota +1 \right)\cos \iota \quad.
\end{equation}
Both parabolas were fitted to the data of all eight bow--shock candidates. The standoff distance with or without
inclination 
and the inclination $\iota$ are given in Table \ref{BSparameters} .\\
A further important parameter characterising a bow shock is the thickness $d$
of the bow--shock
layer. To determine $d$, one plots the variation of the brightness of the
inner boundary, derived from the radial brightness plots,
against the position angle relative to the symmetry axis
resulting in a brightness
profile, shown for \object{HD 57601} in Fig. \ref{BPSkizze}.
The brightness
is given as an emission measure which can be approximated as $EM=n_s^2s$ in
case of a homogeneous
density $n_s$ within the bow--shock layer and the length of the line of sight
$s$ through the layer. Near
the apex of the parabola and at an inclination of $\iota=0\degr$, a segment of 
the layer can be approximated by a spherical shell of  
thickness $d$. In the case of photoionisation equilibrium,
\begin{equation}
\frac{Q}{4\pi}=r^2\alpha_{H\alpha\,A}(T)n_s^2d=r^2\alpha_{H\alpha\,A}(T)EM_{\perp}(d,r)
\end{equation}
is valid. Q is the Lyman continuum flux of the O--star as given for a specific spectral
type by Panagia 
(\cite{Panagia}) and $\alpha_{H\alpha\,A}=5.83\times10^{-14}$\,cm$^3$\,s$^{-1}$ is the
recombination 
coefficient for the H$\alpha$ line at $T=10\,000$\,K derived from Osterbrock (\cite{Osterbrock}). Here, the
emission measure
$EM_{\perp}$ is only valid when looking directly through the bow--shock layer.
The maximum EM we measure is given as
$EM(d,r)=2EM_{\perp}(d,r)\sqrt{ 2\frac{r}{d}+1 }$.
For the parabola of Eq. \ref{burenparabel} the radial distance can be calculated as:
\begin{equation}
\label{EMschraeg}
r(x)=\sqrt{x^2+\left( \frac{x^2}{3R_0}-R_0\right)^2}
\end{equation}
and the polar angle $\theta=\arctan\frac{x}{z}+90\degr$ is given by:
\begin{equation}
x(\theta)=\frac{3}{2}a\tan\theta\pm\sqrt{\frac{3}{2}a^2\tan^2\theta+3a^2}\quad .
\end{equation}
This together leads to a brightness profile of:
\begin{equation}
\label{BP}
EM(d,\theta)=\frac {Q \sqrt{\frac {2r(\theta)} {d}+1} } {2\alpha_{H\alpha\,A}\pi
r^2(\theta)}
\end{equation}
that has to be fitted to the points of the brightness profile, after one has subtracted a
typical 
value of the emission measure of the background measured at the edge of the
images.
In addition, an absorption correction has to be applied as well as
a transformation of angular to linear distances using the
distances of the O--stars, both calculated in Sect. 5. The function of Eq.
\ref{BP} was
fitted and the results are given in
Table \ref{BSparameters} . In the case of \object{HD 149757}, the determination of $d$ was
impossible due to the saturation
of the O--star contaminating the edge of the bow shock
(see Fig. \ref{HalphaBilder1}).

   \begin{table}
      \caption[]{The symmetry axis determined with the IRAS 60\,$\mu$m excess maps ($\theta_{\rm p}$), 
H$\alpha$--images ($\theta_{\rm s}$), and derived from the astrometric data ($\theta_{\rm a}$)
are given and compared using the deviation $\Delta\theta=\theta_{\rm a}-\theta_{\rm s}$. All angles are given in deg.}
         \label{compairsymaxis}
\vspace*{-0.5cm}
     $$ 
         \begin{array}{p{0.2\linewidth} l r@{\pm}l r@{\pm}l r@{\pm}l}
            \hline
            \noalign{\smallskip}
            star& \theta_p& \multicolumn{2}{c}{\theta_\mathrm{s}}& \multicolumn{2}{c}{\theta_\mathrm{a}}& \multicolumn{2}{c}{\Delta\theta}\\
            \noalign{\smallskip}
            \hline
            \noalign{\smallskip}
 HD 24431&   160&      226.0&3.1& 262&38& 36&41\\  
 HD 48099&   340&      301.4&5.3&  71&15& 230&20\\  
 HD 57061&   \cdots& 348.2&2.0& 116.0&6.5& 232.2&8.5\\  
 HD 92206&   166&      112.1&4.1& 147&19& 35&23\\
 HD 135240&  145&      250.3&1.7& 243.7&6.5& 6.3&8.2\\  
 HD 149757&  290&      115.9&1.7&  62.8&1.6& 52.9&3.3\\     
 HD 158186&  \cdots& 306.7&1.7& 317&26& 10&28\\ 
 HD 17505&   154&      183.9&3.8& 207&41& 23&45\\ 
            \noalign{\smallskip}
            \hline
         \end{array}
     $$ 
   \end{table}
\begin{figure*}
\centering
   \includegraphics[angle=0,width=8.9cm]{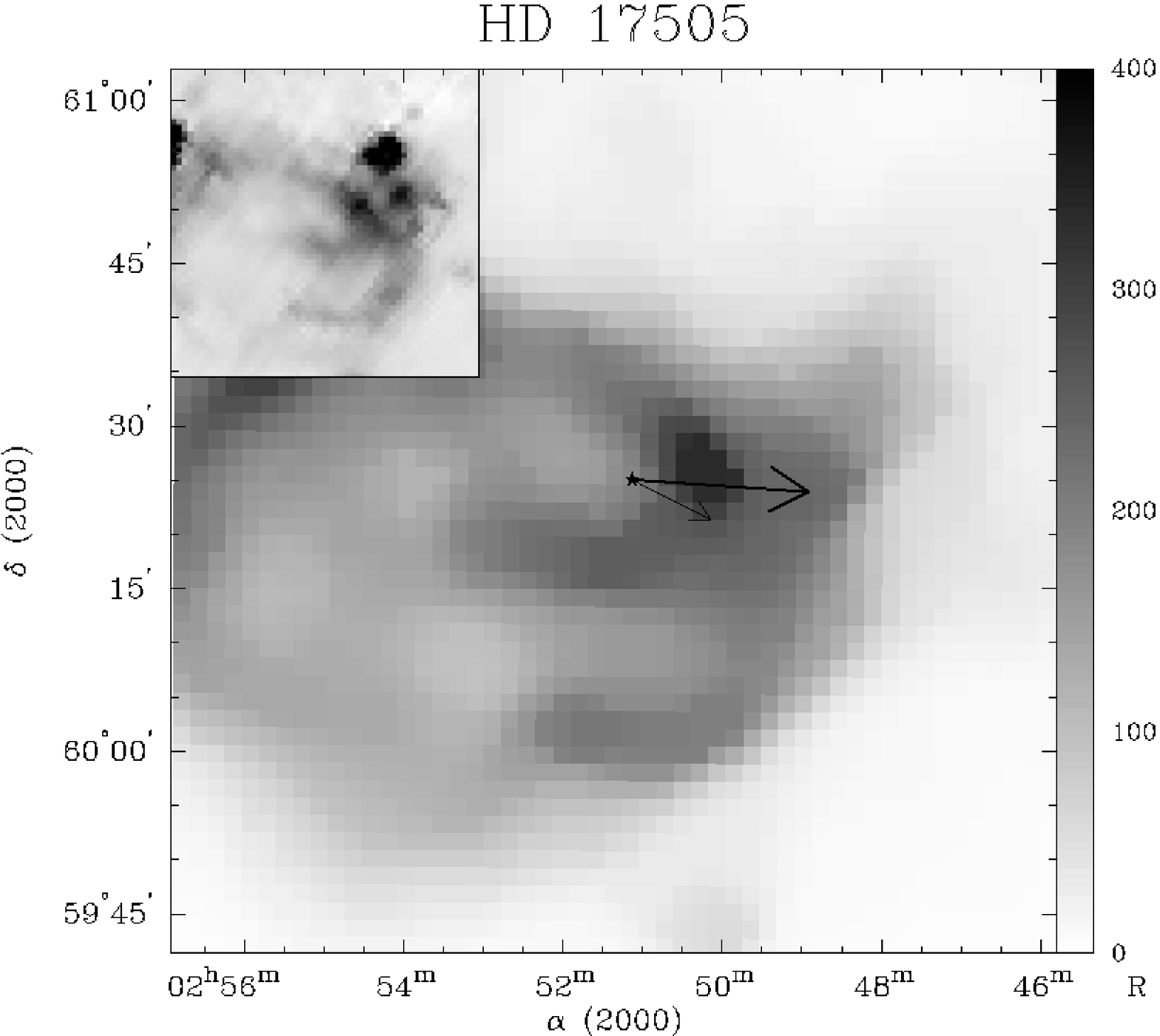}
   \includegraphics[angle=0,width=8.9cm]{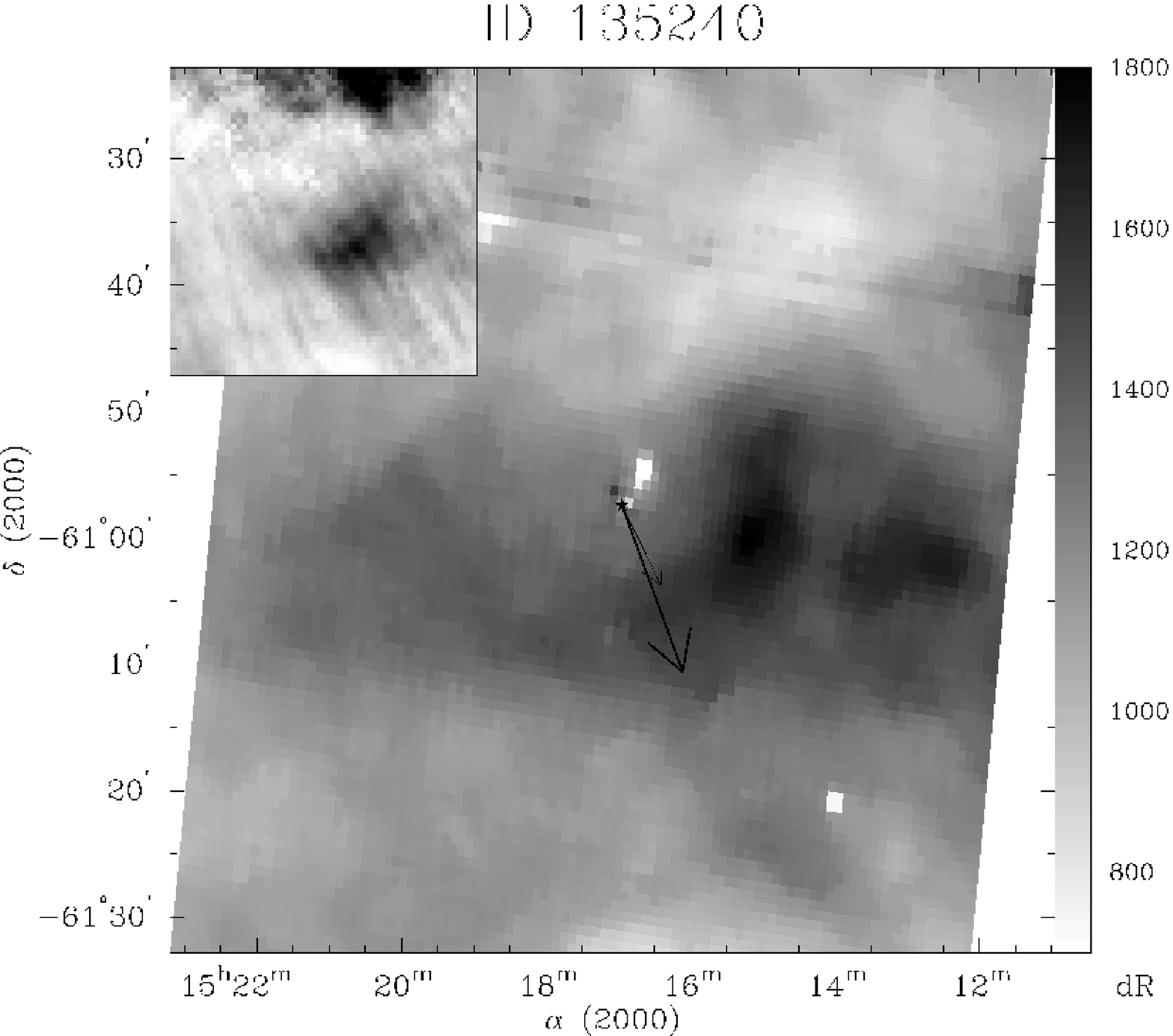}
   \includegraphics[angle=0,width=8.9cm]{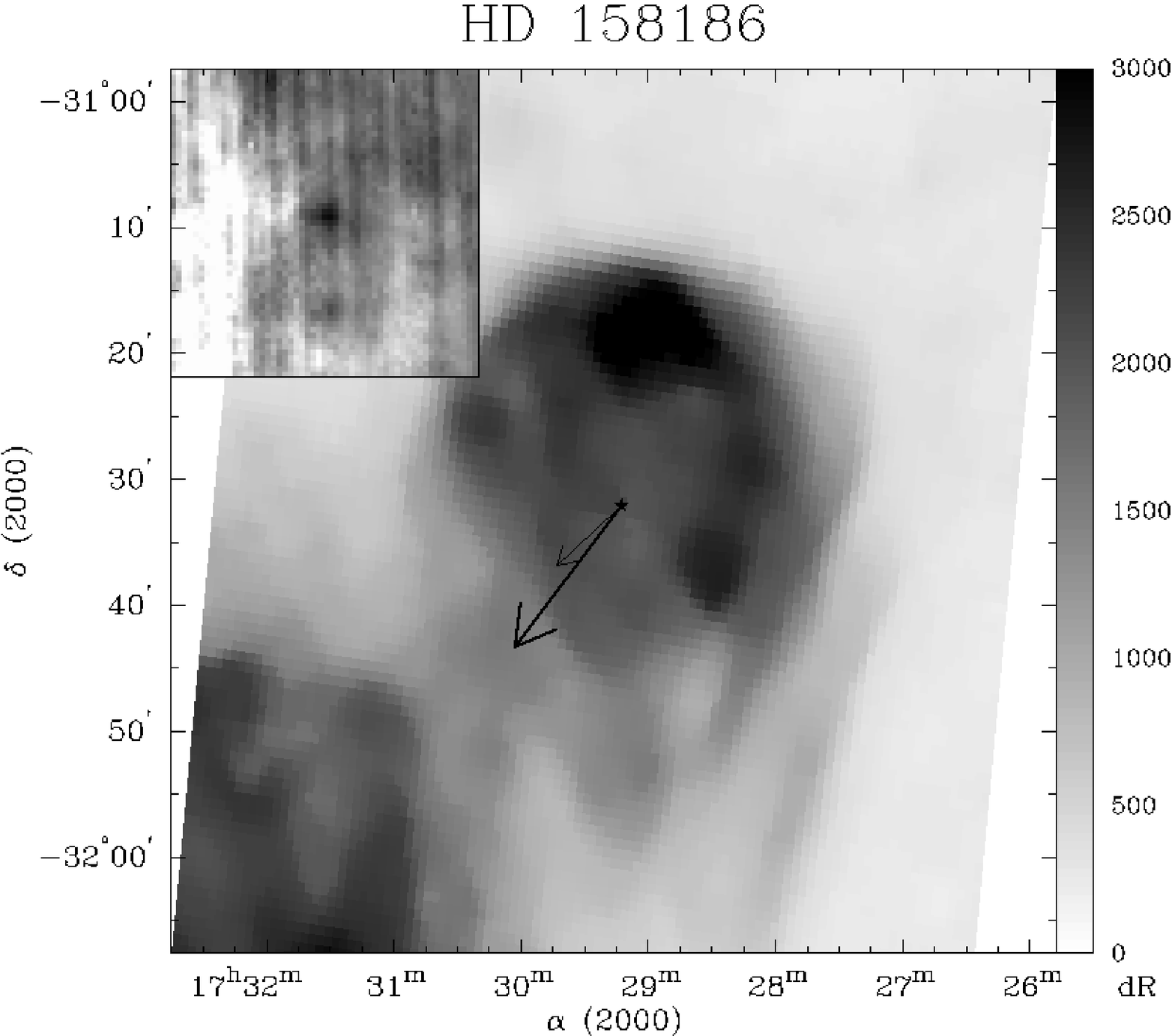}
   \includegraphics[angle=0,width=8.9cm]{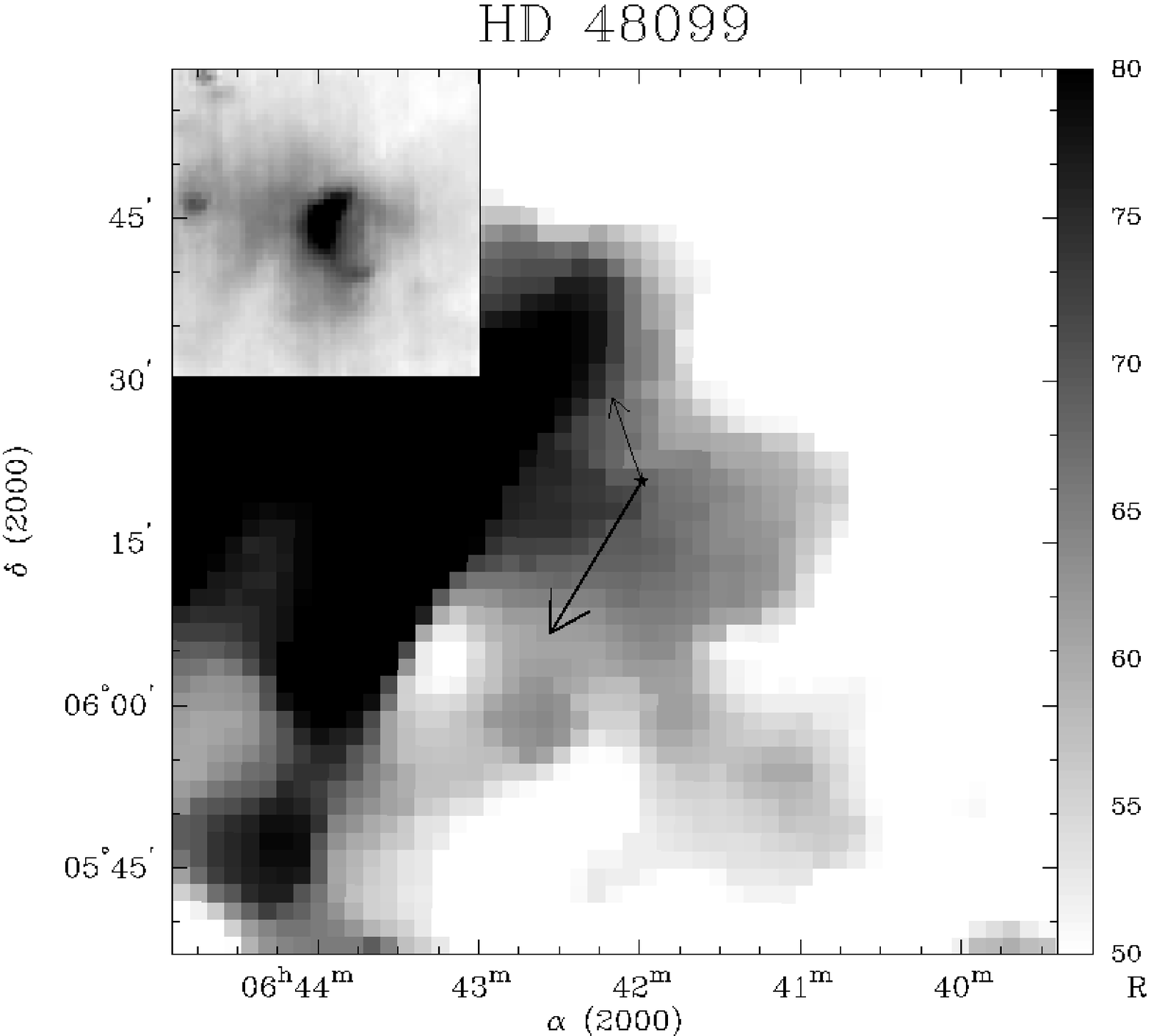}
   \caption{Median filtered H$\alpha$ images of the last four
            bow--shock candidates with their IRAS 60\,$\mu$m excess images,
            symbols are defined as in Fig. \ref{HalphaBilder1}.}
     \label{HalphaBilder2}
\end{figure*}
\section{Results}
It can be seen in Eq. \ref{burenparabel}, that the standoff distance $R_0$ is
the main parameter determining the structure of a bow shock. The
distance of the bow--shock layer at the apex of the parabola, which
is the standoff distance, is determined in the frame of reference of the central star
by the balance of the ram pressure of the
stellar wind and that of the moving ISM. The exact formula of
$R_0$ is given by Wilkin (\cite{Wilkin})
\begin{equation}
  R_0=\sqrt{ \frac{\dot{M}_w V_w}{4 \pi n_{SOD,0} V^2_\ast} }\quad .
\end{equation} 
To further derive the density of the surrounding ISM $n_{SOD,0}$, the mass loss 
rate $\dot{M}_w$ and the asymptotic velocity $V_w$ were taken from 
the literature (see footnotes in Table \ref{ISMparameters}). If non were present, the relations of
van Buren (\cite{Buren83}) and van Buren (\cite{Buren85}) 
based on
the luminosity and effective temperature as given by Panagia (\cite{Panagia})
for the appropriate spectral class
were used to calculate the missing parameters.\\
Due to the large errors of the standoff distance when taking the inclination into
account, only uninclined standoff distances were used to derive the ISM density.
Therefore, all derived ISM densities summarised in
Table \ref{ISMparameters} are lower limits.\\
The temperature of the surrounding ISM is derived due to the fact
that a bow shock can only be created when the central star has 
supersonic velocity $\left ( M=\frac{V_\ast}{C_s} \ge 1\right )$ 
with respect to the surrounding ISM. The
local sound velocity within a neutral medium is
\begin{equation}
\label{sonicvelocity}
  C_s=\sqrt{ \frac{\gamma k T}{\mu n_{SOD,0}} }\quad ,
\end{equation} 
whereby $k$ is the Boltzmann--coefficient and the
adiabatic exponent is assumed as $\gamma=\frac{5}{3}$, because the mean 
mass per particle and proton mass is $\mu=0.61$, when the
Helium fraction is 0.1. As any velocity
higher than sound speed can create a bow shock, the derived temperature 
is only a maximal one and also given in Table \ref{ISMparameters}.\\
As one has derived the thickness $d$, measured the maximum EM of 
the bow--shock layer, and therefore $EM_\perp$, the density of the layer can be derived using the definition 
of the EM$=n_s d$.
The surface density at the apex (cf. Wilkin, \cite{Wilkin})
\begin{equation}
  \sigma=n_s d = \frac{3}{4}R_0n_{LT,0} \left ( 1 + \frac{V_\ast}{V_w}\right )^2
\end{equation}
then leads to an alternative calculation of the
ISM density $n_{LT,0}$ (Table \ref{ISMparameters}).
\section{Discussion}
The results are discussed in three parts. First, the individual characteristics
of the different candidates are mentioned and the problems encountered while analysing
them. Second,
the quality of the developed methods to detect bow shocks and determine their parameters 
are discussed. And thirdly, the sample and its characteristics as a whole
are described and elucidated.
   \begin{table}
      \caption[]{The ISM parameters $n_{SOD,0}$ and $T_{max}$ as derived from the 
                 standoff distance ($R_0$), compared to $n_{LT,0}$ 
                 as derived from bow--shock layer density ($n_s$), using mass loss rates and 
                 stellar wind velocities taken from the literature
                 as cited in the footnotes.}
         \label{ISMparameters}
\vspace*{-0.5cm}
\hspace*{5cm}
     $$ 
         \begin{array}{p{0.2\linewidth} r@{\pm}l r@{\pm}l l r@{\pm}l r@{\pm}l}
            \hline
            \noalign{\smallskip}
            star& \multicolumn{2}{c}{R_0} &\multicolumn{2}{c}{n_{SOD,0}}& T_{max}&
            \multicolumn{2}{c}{n_s} &\multicolumn{2}{c}{n_{LT,0}}\\
            & \multicolumn{2}{c}{[{\rm pc}]} &\multicolumn{2}{c}{[{\rm cm}^{-3}]}& [10^3\,{\rm K}]&
            \multicolumn{2}{c}{[10^3\,{\rm cm}^{-3}]} &\multicolumn{2}{c}{[{\rm cm}^{-3}]}\\
            \noalign{\smallskip}
            \hline
            \noalign{\smallskip}
 HD 24431$^{\mathrm{ad}}$&  1.89&0.05&  3.3&0.2&      6.37&  3.91&0.34&   2.93&0.70         \\  
 HD 48099$^{\mathrm{bd}}$&  2.26&0.10& 0.18&0.02&    62.9 &  4.8&2.1&     1.5&1.0            \\  
 HD 57061$^{\mathrm{cd}}$&  7.56&0.13&   0.07&0.002&   138  &  4.28&0.76&   12.4&5.4          \\  
 HD 92206$^{\mathrm{aa}}$&  3.67&0.19&  0.007&0.001&  7260 &  4810&340&    5.2&1.8^{1}\\
 HD 135240$^{\mathrm{bd}}$& 3.50&0.08&  0.21&0.01&  46.8&  4.56&0.95&   1.43&0.50         \\  
 HD 149757$^{\mathrm{bd}}$& 0.45&0.02&  1.5&0.1&      31.8&  \multicolumn{2}{c}{\cdots}&  \multicolumn{2}{c}{\cdots}\\     
 HD 158186$^{\mathrm{aa}}$& 1.90&0.05&  2.0&0.1&      8.55& 7.3&3.2&     2.96&1.5      \\ 
 HD 17505$^{\mathrm{bd}}$&  1.20&0.04&  21&1&         29.2& 23.6&1.9&     2.47&0.50     \\ 
            \noalign{\smallskip}
            \hline
         \end{array}
     $$ 
\begin{list}{}{}
\item[$^{\mathrm{1}}$]  ISM density of \object{HD 92206} is given in $10^{-3}$\,cm$^{-3}$
\item[$^{\mathrm{a}}$] van Buren (\cite{Buren83}) and van Buren (\cite{Buren85})
\item[$^{\mathrm{b}}$] Howarth \& Prinja (\cite{howarth89})
\item[$^{\mathrm{c}}$] Lamers \& Leitherer (\cite{lamers})
\item[$^{\mathrm{d}}$] Howarth et al. (\cite{howarth97})   
\end{list}
\vspace*{-0.5cm}
   \end{table}
\subsection{Candidates}
As can be seen in Fig. \ref{HalphaBilder2} and Table \ref{compairsymaxis} the symmetry axis of the bow--shock candidate around \object{HD 48099} deviates 
greatly from the proper motion direction. Due to the neighbouring ISM cloud which
dominates the bow--shock nebula in the northeastern side, the axis
is located along the main axis of the neighbouring ISM cloud.\\
In case of \object{HD 149757} the saturation of the central star causes 'bleeding' through the
bow--shock nebula (see Fig. \ref{HalphaBilder1}) and leads to an erroneous direction of the symmetry axis (see Table \ref{compairsymaxis}).
As before the
deviation is in the expected direction, opposite to the defective region.\\
For \object{HD 57061} no explanation can be found for the deviation of the symmetry axis.
The nebula shows typical properties of a bow--shock nebula like its limb brightening (Fig. \ref{BPplot}),
as well as the double hump feature at the apex, seen in Fig. \ref{HalphaBilder1}, predicted by Mac Low et al. (\cite{MacLow}) for
bow shocks inclined like \object{HD 57061}. This could be a hint to a possible misdirection of the astrometrically
determined proper--motion direction of \object{HD 57061}.\\
The bow shock around \object{HD 92206} could not be resolved completely.
Therefore the derived ISM parameters deviate
from the expected values for the Warm Ionised Medium (WIM).\\
Comparing the symmetry axis $\theta_s$ from the H$\alpha$ images
with those derived from the astrometric data $\theta_a$ in Table \ref{compairsymaxis}
it can be said, that they coincide within their errors. Only the above mentioned
cases show a significant misalignment. The same can be said, when comparing
the ISM parameters with expected values for the WIM.
All in all the case of the eight candidates being bow shocks can be significantly strengthened.\\
As for the spatial velocity noted in Table \ref{photo_astro_data}, the following can be said:
the large errors of the velocities of \object{HD 149757} and \object{HD17505}
are sufficient to classify them as runaways. 
As for \object{HD 149757}, it is a long known runaway star (cf. Hoogerwerfer et al. \cite{Hoogerwerfer}).
The determined spacial velocities of \object{HD 24431} and \object{HD 158186} are much lower than the runaway limit, thus the runaway character is uncertain.
However the nebulae detected are most certainly 
bow--shock nebulae. The remaining four candidates have typical velocities for
a runaway OB--star.
\label{Methods}
The radial brightness plot derived using the H$\alpha$ images only shows the expected profile of a
bow shock in case of a few position angles. All profiles only show the brightening 
of the nebula toward its limb, due to the strong background emission.
These profiles are sufficient to determine the position of the inner boundary
of the bow--shock nebula, but are useless as a sole indicator for a bow shock.\\
\begin{figure}
\vspace*{0.5cm}
\centering
   \includegraphics[width=8cm]{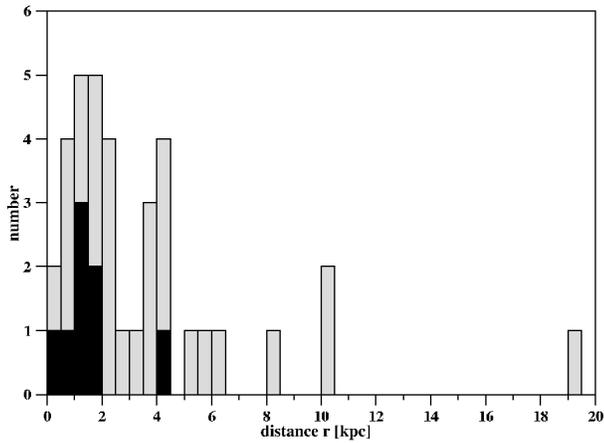}
   \caption{The stellar distance histogram of the complete H$\alpha$ sample. The bow--shock detections
            are plotted in black.}
      \label{BShistogram_r}
\end{figure}
\begin{figure}
\centering
   \includegraphics[width=8cm]{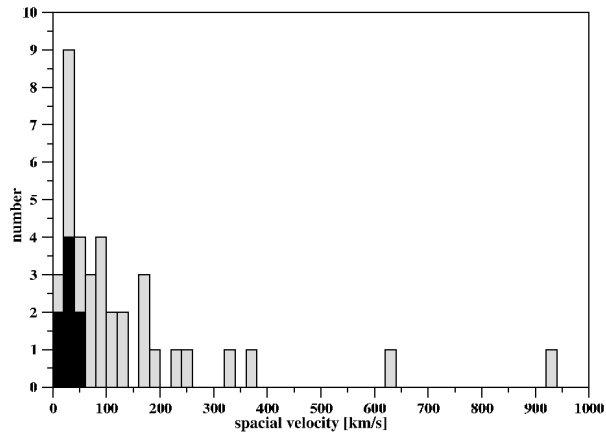}
   \caption{The stellar space velocity histogram of the complete H$\alpha$ sample, as in Fig. \ref{BShistogram_r}.}
      \label{BShistogram_v}
\end{figure}
The best indicator is the correlation of proper--motion and symmetry--axis direction.
Deriving radial distance profiles and their symmetry--axis have been shown to be robust against bright background emission,
through the use of radial brightness plot, as mentioned above. The deviations from the expected profile can easily 
be analysed and put into context with the surrounding ISM, for example when encountering an ISM gradient.\\
Comparing ISM parameters derived from the bow--shock structure with expected WIM parameters
is also a promising method of clarifying the bow--shock character of the nebula. Only the inclination--free
fit of the bow shock leads to a stable fit.
The determination of the standoff distance with respect to the inclination leads to faulty results,
as all candidates seem to either have 
inclination of nearly 0\degr~or 90\degr. Such a high fraction of 
highly--inclined and non--inclined bow shocks is quite
improbable. Especially as the inclination within the used sample is not only  uniformly distributed, but
biased by a preselection of visible bow shocks, such will have low inclinations improving their visibility.\\
Deriving ISM parameters using the determined thickness of the bow--shock layer and
comparing them with the WIM, is also a good method of
clarifying a nebula as a bow shock. The thickness is more difficult to determine, being not
directly measurable. It is derived by fitting an appropriate brightness profile, which results from a crude
model of the brightness distribution along the bow--shock layer. The density of the layer has been taken
as constant, which is only correct in the vicinity of the apex, see Wilkin (\cite{Wilkin}).
Due to the changing angle between bow--shock layer and ISM ram pressure, the force bounding
the layer will also change, resulting in a change of thickness. As the brightness profile fit was only done
near the apex of the parabola, these effects are not as dominant as the errors made when flux correcting.
The correction of the background emission is done by approximating a constant emission, leading
to great uncertainties of the derived thickness. However the ISM parameters derived are
within their errors comparable to the ISM parameters derived from the standoff distance.\\
The remaining problem and limit of all methods is their
ability to discern bow shocks from
stellar bubbles. As demonstrated in Fig. \ref{RBPSkizze} the radial brightness plot shows a clear
limb brightening in both cases. Only the difference in the overall geometry 
(axi--symmetric in the former or spherically symmetric in the latter) enables us to find the bow shock.
But, when taking the surrounding ISM having a density gradient, the structure
of a stellar bubble described by Castor et al. (\cite{castor}) is altered.
The resulting stellar bubble becomes ellipsoidal and the part at the
high--density end will be brighter or the only visible part. Such a deformed stellar bubble
can hardly be discerned from a bow shock, except for its symmetry axis lying
in the direction of the density gradient. The most noticeable difference is the
velocity of the matter in the layer for both cases. In the case of a
stellar bubble, the matter moves radially away from the central star,
whereas the matter in the bow--shock layer moves along the layer and
approximately tangential to the star. Hence, only velocity information
of the matter within the layer, gained either by spectroscopy or
velocity charts within \ion{H}{i} can solve this last ambiguity
(see e.g. Brown \& Bomans \cite{Brown}).
\begin{figure}
\centering
   \includegraphics[width=9cm]{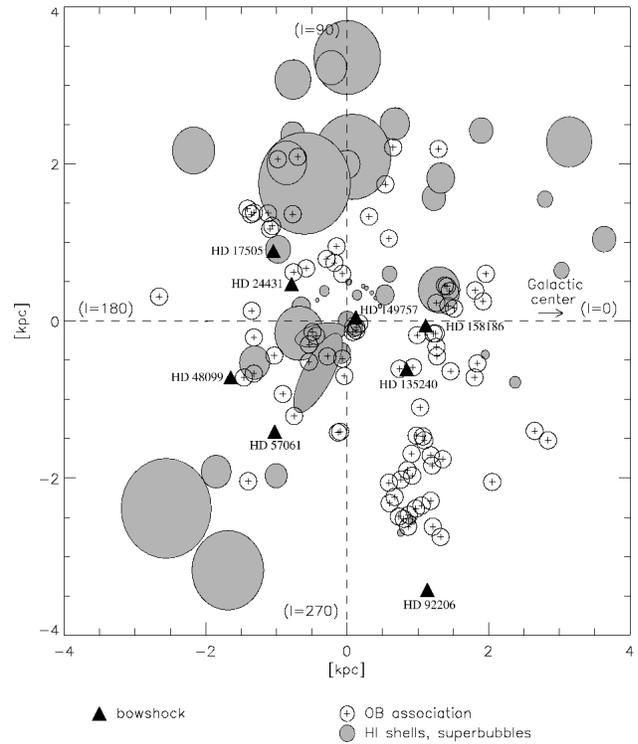}
   \caption{The distribution of the bow shocks in context to superbubbles
            and OB--associations, as published in Huthoff \& Kaper (\cite{huthoff}).}
      \label{BSverteilung}
\end{figure}
\subsection{Sample}
The distance histogram containing all eight detections is compared to that
containing 22 non--detections, without taking into account multiplicity and absorption. The 
resulting histogram, shown in Fig. \ref{BShistogram_r}, reveals that most detections lie within a distance of 2\,kpc. This cannot be a 
result of surface brightness, as it would stay constant with respect to varying distances.
The bow shock, or rather the standoff distance, will vary in size when the distance changes. For example \object{HD 92206}
is the most distant detection and hence is hardly resolved. Similar
results are gained when analysing the velocity histogram in Fig. \ref{BShistogram_v} for the sample. Only bow shocks
around slow central stars can be resolved. One can say this sample is
complete within 2\,kpc and up to velocities of 60\,km\,s$^{-1}$.\\
As described in appendix \ref{Photometric}, the sample contains only one single system. As for the DES and BSS, only 50\% of 
the created runaways should be multiple systems. Whether more bow shocks are created by multiple
systems as for single star systems, or if the fractions of multiples determined from both scenarios
is wrong, cannot be stated on the basis of only eight bow shocks.\\
The location of the eight bow shocks within the galactic plane in Fig. \ref{BSverteilung} shows that they are not located
inside any superbubble. This is to be expected, as the sonic velocity (Eq. \ref{sonicvelocity}, with $T \sim 10^6$\,K) therein would be
to high to create a bow shock. Only \object{HD 17505} seems to lie in the superbubble inside \object{GS 137--27--17},
which is only an effect of projection. \object{HD 17505} having a $b=-0.9$\degr~lies well above the upper 
surface of the bubble at $b=0$\degr. For the three bow shocks with $270\degr\le l\le 0\degr$ no
data for superbubbles exists.\\
The typical lifetime of an OB--star with its runaway velocity does not permit it to put a
large distance between itself and the OB--association as a probable origin. As shown in
Fig. \ref{BSverteilung} most bow shocks are still near OB--associations, except \object{HD 92206}, \object{HD 57061} and \object{HD 17505}. 
The first two, are the fastest stars within the sample, enabling them to move a greater distance, and \object{HD 92206} is also so far 
away, that OB--associations can only be determined with great difficulty at such a distance.
\object{HD 17505} is in the vicinity of
a star formation region noted in Carpenter et al. (\cite{Carpenter}) not shown in Fig. \ref{BSverteilung}.
\section{Conclusion}
The search for bow shocks using the H$\alpha$--sample of SHASSA and VTSS
yielded eight detections (\object{HD 17505}, \object{HD 24430}, \object{HD 48099}, \object{HD 57061}, \object{HD 92206}, \object{HD 135240}, \object{HD 149757}, and \object{HD 158186}) from a total
of 30 candidates already observed (seven candidates are missing in VTSS),
derived from the sample VB used for their search within the IRAS allsky survey.\\
The best indicator to detect bow shocks
within H$\alpha$ images was the correlation between the direction
of the proper--motion and the symmetry axis, determined using radial distance profiles, which are
not sensitive to bright backgrounds.
The other methods can be used for further verification, when in doubt (as for HD 57061).\\
The detected bow shocks could be successfully used to determine ISM parameters.
This was done, either using the standoff distance or the brightness profile. As the sample 
is only complete up to a distance of 2\,kpc and no bow shocks can be found inside
a superbubble, the derived values of the 
density ($\sim 1$\,cm$^{-3}$) and the maximal temperature ($\sim 10^4$\,K) fit
well to the picture of the WIM (e.g. Shull \cite{Shull}). 
Both features justify the more or less constant ISM density and temperature.
Regions of other ISM composition and therefore other parameters, e.g. the regions inside
the Small and Large Magellanic Cloud,
are too far away and cannot be analysed using the medium resolution
of the H$\alpha$--allsky surveys.  
All in all, bow shocks around OB--runaway stars are ideal probes of the ISM,
as are neutron stars (cf. Chatterje \& Cordes \cite{chatterjee}). 
They also demonstrate the qualitative picture of the neighbouring
ISM is apparently correct.\\
First steps to clarify the last ambiguity of bubble or bow shock using spectroscopy
or velocity charts (see Sect. \ref{Methods}) were taken for the eight candidates:
Using data from the IUE-archive, absorption profiles of \ion{N}{v}, \ion{Si}{iv}, and \ion{C}{iv}
were detected in the case of \object{HD 48099}, \object{HD 57060},
and \object{HD135240} showing an excess compared to surrounding O--stars
and low temperatures of $\sim$15\,000\,K. As stellar winds are too thin to contribute
large amounts of these ions and the temperatures are
too low to radiatively excite them, this confirms again these objects
as shock fronts of bow shocks.
More important, these lines were used to measure the velocities, which fit to the proposed values
of the bow--shock layer. 
Additionally, \ion{H}{i} velocity data of \object{HD 17505} 
from the Canadian Galactic Plane Survey was used to create velocity charts,
which showed a velocity distribution as expected in the case of a bow--shock layer.
However no data could be found in the literature concerning the other candidates.
Hence further spectroscopic analysis of these objects is still needed to
make use of the full potential of the current sample as ISM probes.

\begin{acknowledgements}
We are grateful to L. Kaper and F. Huthoff for permission to use their results.
Also we want to thank N. Bennert and K. Weis for their careful reading and
useful suggestions. The use of the SIMBAD database, operated by CDS, Strasbourg, France,
is acknowledged. This research has also made use of the NASA\/ IPAC Infrared
 Science Archive, which is operated by the Jet Propulsion
Laboratory, California Institute of Technology, under
contract with the National Aeronautics and Space
Administration. DB thanks R.\,J. Dettmar for support. DJB acknowledges
the German \emph{Deut\-sche For\-schungs\-ge\-mein\-schaft, DFG\/} project
SFB 591 'Universal Behavior of non--equilibrium Plasmas'.

\end{acknowledgements}

\appendix

\section{Photometric Data}
\label{Photometric}
{\bf \object{HD 135240}:} Penny et al. (\cite{Penny}) have determined \object{HD
135240}
as a triple system and specified the first component as a O7 III--V, the second
as a O9.5 V and the last one as a B0.5 V star. They also measured the UV flux
ratios
for the different components as $\frac{F_{UV2}}{F_{UV3}}=0.239$ and 
$\frac{F_{UV3}}{F_{UV1}}=0.179$, so that the magnitudes of the single
components
could be determined from the total magnitudes given in the SIMBAD--database.\\
{\bf \object{HD 57061}:} For photometric data, the SIMBAD--database was used and
the results of Stickland et al. (\cite{Stickland}) of the star being a
quintuple system. They could resolve one component as a O9 II star. The other
consists of two double systems resulting in a total spectral class of B0.5 V.
The single O--star is the dominant star with a ten--times higher flux than the
second double binary.\\
{\bf \object{HD 17505}:} Fabricius \& Makarov (\cite{Fabricius}) note this star
as being a double or triple system and could measure the magnitudes of the
brightest component using the Tycho--filtersystem. The spectral class of O9.5 V
(Garmany et al. \cite{Garmany})
can only be determined for the complete system.\\
{\bf \object{HD 158186}:} This object was detected as being a variable of
the Algol--type by Adelmann et al. (\cite{Adelmann}), therefore \object{HD
158186}
has to be at least a double system. As the system is not resolved, 
the star was treated as a single star with photometric data
from the SIMBAD--database and the total spectral class of O9.5 V given by 
Buscombe (\cite{Buscomb}).\\
{\bf \object{HD 24431}:} Fabricius \& Makarov (\cite{Fabricius}) also analysed
this
star and classified it at least as a double system and were able to measure the
magnitude of two components. As in the case of \object{HD 17505} they used the
Tycho--filtersystem. Reed (\cite{Reed}) states the spectral class as a O9
IV--V.\\
{\bf \object{HD 92206}:} The SIMBAD--database notes this star as a double
system,
but can only give the total magnitude of the system. Therefore it is treated
as
a single star. Reed (\cite{Reed}) could only measure the spectral class of O6,
and was not
able to determine the luminosity class.\\
{\bf \object{HD 149757}:} Here the SIMBAD--database notes the star as a single
system
and its magnitude. The spectral class was measured by Garmany (\cite{Garmany})
and is
O9 V. Being the nearest star of the eight candidates, the distance
determined
by Hipparcos could be determined. Both distances are within reasonable agreement
with
respect to their errors, verifying the spectral parallax results.
For consistency, we use the distance
determined
by spectral parallax.\\
{\bf \object{HD 48099}:} This star is given as a binary system by Stickland et
al. 
(\cite{Stickland}). They could measure the UV--flux ratios
for the components as $\frac{F_{UV1}}{F_{UV2}}=1.8$. Garmany
(\cite{Garmany})
could only determine the total spectral class to O7 V.\\

\end{document}